\documentclass[11pt,a4paper]{article}
\usepackage{psfig,amssymb,alltt}
\usepackage{latexsym}
\usepackage{graphicx}
\usepackage{psfig}
\usepackage{amsmath}

\psfigurepath{./PS}

\newcommand{\be}{\begin{eqnarray}}
\newcommand{\ee}{\end{eqnarray}}
\newcommand{\goto}{\rightarrow}
\newcommand{\bitem}{\begin{itemize}}
\newcommand{\eitem}{\end{itemize}}
\newcommand{\bR}{{\bf R}}

\newcommand{\cR}{{\cal R}}

\newcommand{\wave}{{\cal W}}

\title{Analysis of the Spatial Distribution of Galaxies by Multiscale Methods}
\author{  J-L. Starck  \\
        DAPNIA/SEDI-SAP, Service d'Astrophysique, \\
    CEA-Saclay, 91191 Gif-sur-Yvette, France   \\ [12pt]
    V.J.\,Mart\'\i nez \\
   Observatori Astron\`omic de la Universitat de \\
   Val\`encia, Edifici d'Instituts de Paterna, Apartat de Correus \\
   22085, 46071 Val\`encia, Spain\\ [12pt]
      D. L. Donoho, O. Levi \\
     Department of Statistic, Stanford University, \\
          Sequoia Hall, Stanford, CA 94305, USA \\ [12pt]
      P. Querre   \\
      DAPNIA/SEDI-SAP, Service d'Astrophysique, \\
      CEA-Saclay, 91191 Gif-sur-Yvette, France \\ [12pt]
      E. Saar     \\
      Tartu Observatory, Toravere, 61602 Estonia
       \\
      \vspace{1cm}}

\begin{document}
\maketitle

\begin{abstract}
 Galaxies are arranged in interconnected walls and filaments
forming a cosmic web encompassing huge, nearly empty, regions
between the structures. Many statistical methods have been proposed
in the past in order to describe the galaxy distribution and
discriminate the different cosmological models. We present in this
paper results relative to the use of new statistical tools using the
3D isotropic undecimated wavelet transform, the 3D ridgelet transform and the 3D beamlet
transform. We show that such multiscale methods produce a new way to
measure in a coherent and statistically reliable way the degree of
clustering, filamentarity, sheetedness, and voidedness of a dataset.
\end{abstract}

\begin{center}
{\bf Keywords} \\

Galaxy distribution, large scale structures, wavelet, ridgelet, beamlet, multi-scale methods.
\end{center}

\section{Introduction}

Galaxies are not uniformly distributed throughout the universe.
Voids, filaments, clusters, and walls of galaxies can be observed,
and their distribution constraints our cosmological theories.
Therefore we need reliable statistical methods to compare the
observed galaxy distribution with theoretical models and
cosmological simulations.

The standard approach for testing models is to define a point
process which can be characterized by statistical descriptors. This
could be the distribution of galaxies of a specific type in deep
redshift surveys of galaxies (or of clusters of galaxies). In order
to compare models of structure formation, the different distribution
of dark matter particles in N-body simulations could be analyzed as
well, with the same statistics.

The two-point correlation function $\xi(r)$ has been the primary
tool for quantifying large-scale cosmic structure
\cite{cf:peebles80}. Assuming that the galaxy distribution in the
Universe is a realization of a stationary and isotropic random
process, the two-point correlation function can be defined from
the probability $\delta P$ of finding an object within a volume
element $\delta V$ at distance $r$ from a randomly chosen object
or position inside the volume: $ \delta P = n(1 + \xi(r))\delta
V$, where $n$ is the mean density of objects.  The function
$\xi(r)$ measures the clustering properties of objects in a given
volume. It is zero for a uniform random distribution, positive
(respectively, negative) for a more (respectively, less) clustered
distribution. For a hierarchical clustering or fractal process,
$1+\xi(r)$ follows a power-law behavior with exponent $D_2-3$.
Since $\xi(r) \sim r^{-\gamma}$ for the observed galaxy
distribution, the correlation dimension for the range where
$\xi(r) \gg 1$ is $D_2 \simeq 3-\gamma$. The Fourier transform of
the correlation function is the power spectrum. The direct
measurement of the power spectrum from redshift surveys is of
major interest because model predictions are made in terms of the
power spectral density. It seems clear that the real space power
spectrum departs from a single power-law ruling out simple
unbounded fractal models \cite{tegmark04}. The two-point
correlation function can been generalized to the N-point
correlation function \cite{cf:szapudi98,cf:peebles01}, and all the
hierarchy can be related with the physics responsible for the
clustering of matter. Nevertheless they are difficult to measure,
and therefore other related statistical measures have been
introduced as a complement in the statistical description of the
spatial distribution of galaxies \cite{cf:martinezsaar}, such as
the void probability function \cite{lachieze}, the multifractal
approach \cite{martinez90}, the minimal spanning tree
\cite{astro:bhavsar96,astro:krzewina96,astro:doro01}, the
Minkowski functionals \index{Minkowski!functionals} \index{J
function} \cite{cf:mecke94,cf:kerscher00b} or the $J$ function
\cite{cf:lieshout96,kerscherJ} which is defined as the ratio $J(r)
= \frac{1-G(r)}{1-F(r)}$, where $F$ is the distribution function
of the distance $r$ of an arbitrary point in $\bR^3$ to the
nearest object in the catalog, and $G$ is the distribution
function of the distance $r$ of an object to the nearest object.
Wavelets have also been used for analyzing the projected 2D or the
3D galaxy distribution
\cite{xwave:escalera92b,astro:slezak93,martinez93,astro:pagliaro99,cf:kurokawa01}.

The Genus statistic \cite{gott86} measures the topology or the
degree of connectedness of the underlying density field.  Phase
correlations of the density field can be detected by means of this
topological analysis. The Genus is calculated by (i) convolving
the data by a kernel, generally a Gaussian, (ii) setting to zero
all values under a threshold $\nu$ in the obtained distribution,
and (iii) taking the difference $D$ between the number of holes
and the number of isolated regions. The Genus curve $G(\nu)$ is
obtained by varying the threshold level $\nu$. The first step of
the algorithm, the convolution by a Gaussian, may be dramatic for
the description of filaments, which are spread out along all
directions, as it will be shown in next section. The Genus is
related with one of the four Minkowski functionals that describe
well the overall morphology of the galaxy distribution
\cite{varun1}. Minkowski functionals have been used to elaborate
sophisticated tools to measure the filamentarity and planarity of
the distribution by means of shape finders quantities
\cite{varun2}.

The Sloan Digital Sky Survey (Early Data Release) has recently be
analyzed using a 3D Genus Statistics \cite{astro:hikage02} and
results were consistent with that predicted by simulations of a
$\Lambda$-dominated spatially-flat cold dark matter model.

New multiscale methods have recently emerged, the beamlet transform
\cite{cur:donoho_01,cur:donoho02} and the ridgelet transform
\cite{cur:candes99_1}, which allows us to better represent data
containing respectively filaments and sheets, while wavelets
represent well isotropic features (i.e. cluster in 3D). As each of
these three transforms represents perfectly one kind of feature, all
of them are useful and should be used to describe a given catalog.

Section~2 describes the 3D wavelet transform, and how wavelets can
be used for estimating the underlying density. Sections 3 and 4
describes respectively the 3D ridgelet transform and the 3D
beamlet transform. It is shown in section~4 through a set of of
experiments how these three 3D transforms can be combined in order
to describe statistically the distribution of galaxies.

\section{The 3D Wavelet Transform}

\subsection{The Undecimated Isotropic Wavelet Transform}
The wavelet transform of a signal produces, at each scale $j$, a
set of zero-mean coefficient values $\{w_j\}$. Using an algorithm
such as the undecimated isotropic wavelet decomposition
\cite{starck:book98}, this set $\{w_j\}$ has the same number of
pixels as the signal and thus this wavelet transform is a
redundant one. Furthermore, using a wavelet defined as the
difference between the scaling functions of two successive scales
\begin{eqnarray}
\frac{1}{8} \psi(\frac{x}{2}, \frac{y}{2}, \frac{z}{2}) = \phi(x,y,z)
     - \frac{1}{8} \phi(\frac{x}{2},\frac{y}{2},\frac{z}{2}),
\end{eqnarray} the original cube $c=c_{0}$ can be expressed as the
sum of all the wavelet scales and the
 smoothed array $c_{J}$
\begin{eqnarray}
c_{0,x,y,z} = c_{J,x,y,z} + \sum_{j=1}^{J} w_{j,x,y,z} \ ,
\end{eqnarray}
The set $w = \{ w_1, w_2, ..., w_J, c_J \}$, where
$c_J$ is a last smooth array, represents the
wavelet transform of the data. If we denote as $\wave$
the wavelet transform operator
and $N$ the pixel number of $c$, the wavelet transform
$w$ ($w = \wave c$) has $(J+1)N$ pixels (redundancy factor of
$J+1$). The scaling function $\phi$ is generally chosen as a
spline of degree 3, and the 3D implementation is based on three 1D
sets of (separable) convolutions. Like the scaling function
$\phi$, the wavelet function $\psi$ is isotropic (point
symmetric).
More details can be found in \cite{starck:book98,starck:book02}.

For each $a > 0$, $b_1,b_2,b_3 \in \bR^3$ , the {\em wavelet} is defined by

\noindent {\parindent=4cm \indent \(
\begin{array}{ll}
\label{eq:waveletelet}
       \psi_{a,b_1,b_2,b_3}: &\bR^3 \goto \bR \\
       &\psi_{a,b_1,b_2,b_3} (x_1,x_2,x_3) = a^{-1/2} \cdot
       \psi( \frac{x_1-b_1}{a}, \frac{x_2-b_2}{a}, \frac{x_3-b_3}{a})
\end{array} \)
}

Given a function $f \in L^2(\bR^3)$, we define its wavelet
coefficients by:

\noindent {\parindent=4cm \indent\(
\begin{array}{lll}
        \mathcal{W}_f: &\bR^4 \goto \bR \\
               &\mathcal{W}_f(a,b_1,b_2,b_3) = \int \overline{\psi}_{a,b_1,b_2,b_3}
           (\mathbf{x}) f(\mathbf{x}) d\mathbf{x}.
\end{array}\)
}
\begin{figure}[htb]
\centerline{
\hbox{
\psfig{figure=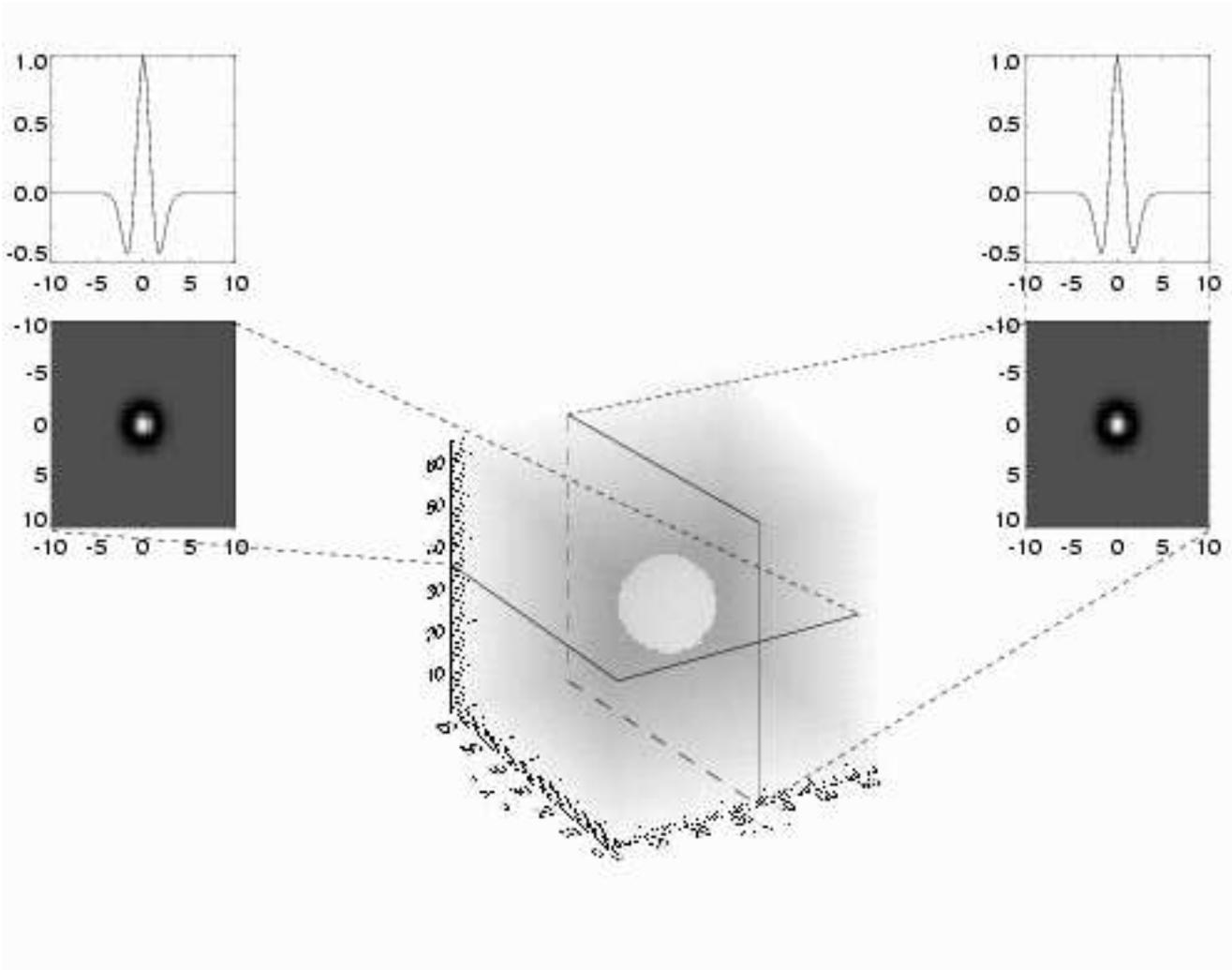}
}}
\caption{Example of wavelet function.}
\label{fig_wave3d}
\end{figure}

Figure~\ref{fig_wave3d} shows an example of 3D wavelet function.

\subsection{3D Galaxy Distribution Filtering}

For the noise model, given that this relates to point pattern clustering,
we have to consider the Poisson noise case. The autoconvolution histogram
method used for X-ray image  \cite{starck:sta98_1} can also be used here.
It consists to calculate numerically the probability distribution function (pdf) of a wavelet $w_{j,x,y,z}$
coefficient with the hypothesis that the galaxies used for obtaining $w_{j,x,y,z}$ are
randomly distributed. The pdf is obtained by autoconvolving $n$ times the histogram of the wavelet function,
$n$ being the number of galaxies which have been used for obtaining $w_{j,x,y,z}$, i.e. the number
of galaxies in a box around $(x,y,x)$, the size of the box depending of the scale $j$.
More details can be found in \cite{starck:sta98_1,starck:book02}.

Once the pdf relative to the coefficient $w_{j,x,y,z}$ is known, we can detect the significant
wavelet coefficients easily. We derive two threshold values
$T^{min}_{j,x,y,z}$ and $T^{max}_{j, x,y,z}$ such that
\begin{eqnarray}
Prob (W < T^{min}_{j,x,y,z}) & = & \epsilon \nonumber \\
Prob (W > T^{max}_{j,x,y,z}) & = & \epsilon
\label{TestHyp}
\end{eqnarray}
$\epsilon$ corresponding to the confidence level,
and the positive (respective negative) wavelet coefficient is significant if it is larger
than $T^{max}_{j,x,y,z}$ (resp. lower than $T^{min}_{j,x,y,z}$).

A simple filtering method would now consist to set to zero (i.e. thresholding) all unsignificant
coefficients and reconstruct the filtered data cube by addition of the different scales.
But when a redundant wavelet transform is used, the result after a simple hard
thresholding can still be improved by iterating \cite{starck:sta95_1}. We want  the
wavelet transform of our solution $\tilde s$ to reproduce the same significant
wavelet coefficients (i.e., coefficients larger than $T_j$). This can
be expressed in the following way:
\begin{eqnarray}
 (\wave \tilde s)_{j,k} = w_{j,k} \mbox{ if  } w_{j,k}   \mbox{ is significant }
\end{eqnarray}
where $w_{j,k}$ are the wavelet coefficients of the input data $s$ at scale $j$ and at position $k=(x,y,z)$.
The relation is not necessarily verified in the case of non-orthogonal
transforms,
and the resulting effect is generally a loss of flux inside the objects.
The residual signal (i.e. $s - \tilde s$) still contains some information
at positions where the objects are.

Denoting $M$ the multiresolution support of $s$ (i.e. $M_{j,k} = 1$ if
$w_{j,k}$ is significant, and 0 otherwise), we want:
\begin{eqnarray*}
 M.\wave \tilde s  = M.\wave s
\end{eqnarray*}
The solution can be obtained by the following Van Cittert
iteration \cite{starck:book98}:
\begin{eqnarray}
 \tilde{s}^{n+1} & = & \tilde{s}^n +  \wave^{-1} (M.\wave s - M.\wave s^n)  \nonumber \\
          & = & \tilde{s}^n +  \wave^{-1} (M.\wave R^n)
\label{eqn_iter_support}
\end{eqnarray}
where $R^n = s - \tilde{s}^n$.

\subsubsection*{Iterative Filtering with a Smoothness Constraint}
\label{sect_filter_regul}
A smoothness constraint can be imposed on the solution.
\begin{equation}
  \label{eq:l1-min}
  \min \|\wave\tilde{s}\|_{\ell_1}, \quad \mbox{subject to} \quad s \in C,
\end{equation}
where $C$ is the set of vectors $\tilde{s}$
which obey the linear constraints
\begin{equation}
\label{eq:constraints}
\left\{  \begin{array}{ll}
  \tilde{s}_k \ge 0, \forall k \\
 \mid  (\wave \tilde{s} - \wave s)_{j,k} \mid \le e_j;
  \end{array}
  \right.
\end{equation}
Here, the second inequality constraint
only concerns the set of significant coefficients,
i.e. those indices which exceed (in absolute value)
a detection threshold $t_j$. Given a
tolerance vector $e = \{e_1, ..., e_j\}$,
we seek a solution whose coefficients
 $(\wave \tilde{s})_{j,k}$, at scale and position where significant
 coefficients were detected,  are within $e_j$
 of the noisy coefficients $(\wave s)_{j,k}$. For example, we can choose
 $e_j =  \sigma_j / 2$. In short,
our constraint guarantees that the reconstruction be smooth but
will take into
account any pattern which is detected as significant by the wavelet transform.

We use an $\ell_1$ penalty on the coefficient sequence because we are
interested in {\em low complexity} reconstructions. There are other
possible choices of complexity penalties; for instance, an alternative
to (\ref{eq:l1-min}) would be
\[
\label{eq:tv-min}
  \min \|\tilde{s}\|_{TV}, \quad \mbox{subject to} \quad s \in C.
\]
where $\|\cdot\|_{TV}$ is the Total Variation norm, i.e.\ the discrete
equivalent of the integral of the Euclidean norm of the gradient.
\index{total variation}
Expression (\ref{eq:l1-min}) can be solved using the method of hybrid
steepest descent (HSD) \cite{wave:yamada01}. HSD consists of building
the sequence
\begin{eqnarray}
 \tilde{s}^{n+1} = P(\tilde{s}^{n}) - \lambda_{n+1} \nabla_J(P(\tilde{s}^{n}));
\end{eqnarray}
here, $P$ is the $\ell_2$ projection operator onto the feasible set
$C$, $\nabla_J$ is the gradient of equation~(\ref{eq:l1-min}), and
$(\lambda_{n})_{n \ge 1}$ is a sequence obeying $(\lambda_{n})_{n\ge
  1} \in [0,1] $ and $\lim_{ n \rightarrow + \infty } \lambda_{n} = 0$.

Unfortunately, the projection operator $P$ is not easily determined
and in practice we use the following proxy; compute $\wave \tilde{s}$
and replace those coefficients which do not obey the constraints $| (\wave
\tilde{s} - \wave s)_{j,k} | \le e_j$
(those which fall outside of the prescribed
interval) by those of $s$; apply the inverse transform.

The filtering algorithm is:
\begin{enumerate}
\baselineskip=0.4truecm
\itemsep=0.1truecm
\item Initialize $L_{\max} = 1$, the number of iterations $N_i$, and
  $\delta_{\lambda} = \frac{L_{\max}}{N_i}$.
\item Estimate the noise standard deviation $\sigma_s$, and set $e_j =
  {\sigma_j \over 2}$ for all $j$.
\item Calculate the transform: $w^{(s)} = \wave s$.
\item Set $\lambda = L_{\max}$, $n = 0$, and $\tilde s^{n}$ to 0.
\item While $\lambda >= 0$ do
\begin{itemize}
\item $u = \tilde s^{n}$.
   \begin{itemize}
     \item Calculate the transform $\alpha = \wave u$.
     \item For all coefficients $\alpha_{j,k}$ do
     \begin{itemize}
     \item Calculate the residual $r_{j,k} = w^{(s)}_{j,k} - \alpha_{j,k}$
      \item if $w^{(s)}_{j,k}$ is significant and $ \mid r_{j,k}
       \mid \ > e_{j}$ then $\alpha_{j,k} = w^{(s)}_{j,k}$
     \item $\alpha_{j,k} = sgn(\alpha_{j,k}) ( \mid \alpha_{j,k} \mid - \lambda \sigma_j)_{+}$.
     \end{itemize}
   \item $u = {\wave}^{-1} \alpha$
  \end{itemize}
\item Threshold negative values in $u$ and $\tilde s^{n+1} = u$.
\item $n = n + 1$, $\lambda = \lambda - \delta_{\lambda} $, and goto 5.
\end{itemize}
\end{enumerate}
In practice, a small number of iterations ($<$10) is enough.
The $(.)_+$ operator means that negative values are set to zero
($(a)_+ = \mbox{MAX}(0,a)$).

\subsubsection*{Experiments}
\begin{figure}[htb]
\centerline{
\hbox{ 
\psfig{figure=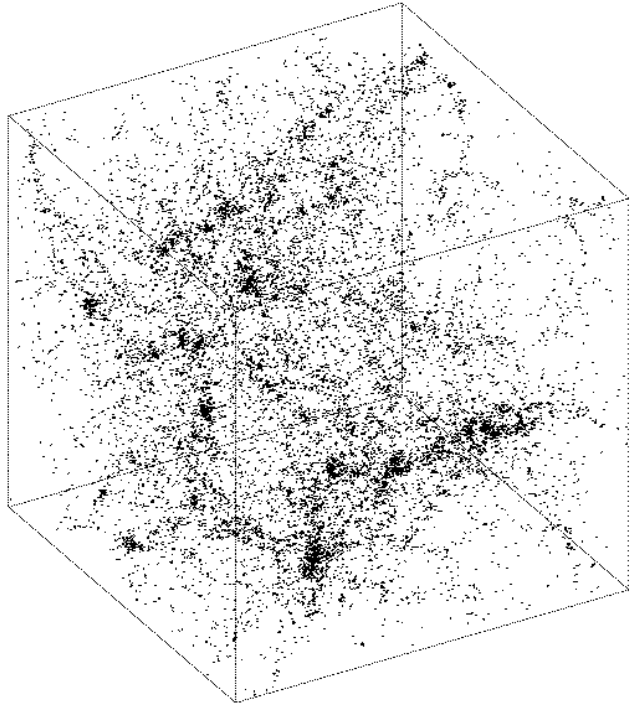,width=10cm,height=10cm}
}}
\caption{Simulated data.}
\label{fig_lss}
\end{figure}

Fig.~\ref{fig_lss} shows a simulation of a $60 h^{-1}$ Mpc box of
universe, made by A.\ Klypin (simulated data at
http://astro.nmsu.edu/$\sim$aklypin/PM/pmcode). It represents the
distribution of dark matter in the present-day universe and each
point is a dark matter halo where visible galaxies are expected to
be located, i.e. the distribution of dark matter halos can be
compared with the distribution of galaxies in catalogs of
galaxies.

\begin{figure}[htb]
\centerline{ 
\resizebox{0.32\textwidth}{!}{\includegraphics*{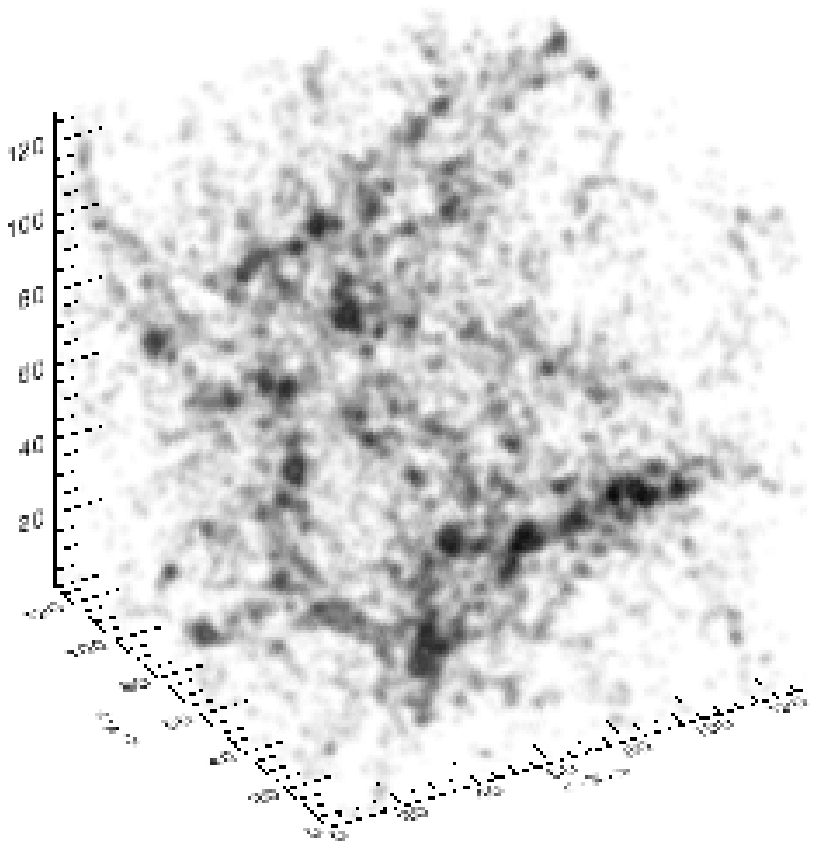}}
\resizebox{0.32\textwidth}{!}{\includegraphics*{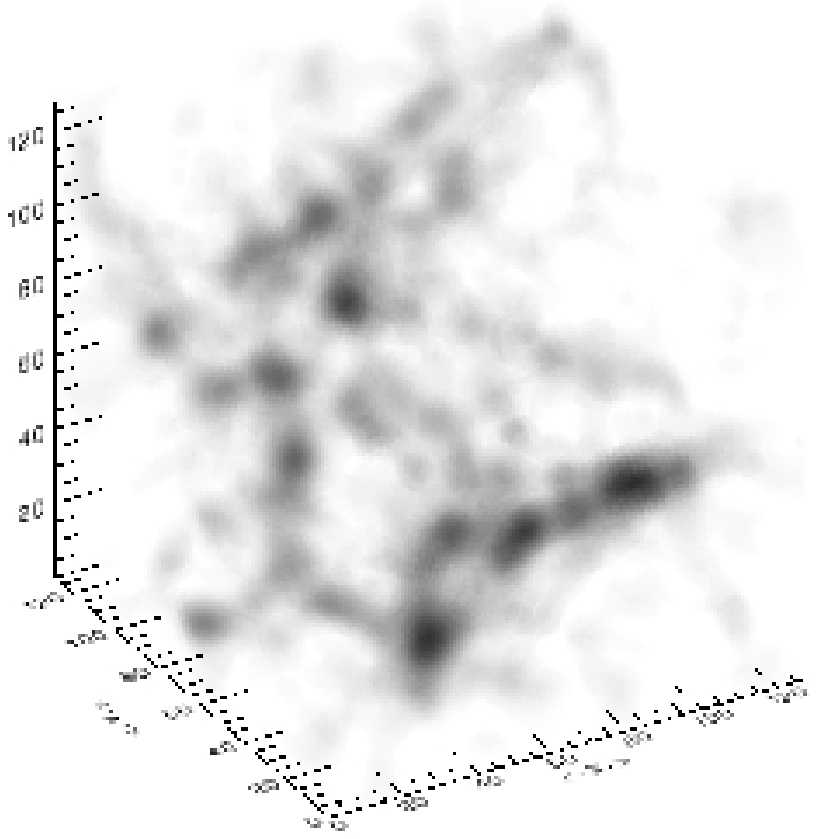}}
\resizebox{0.32\textwidth}{!}{\includegraphics*{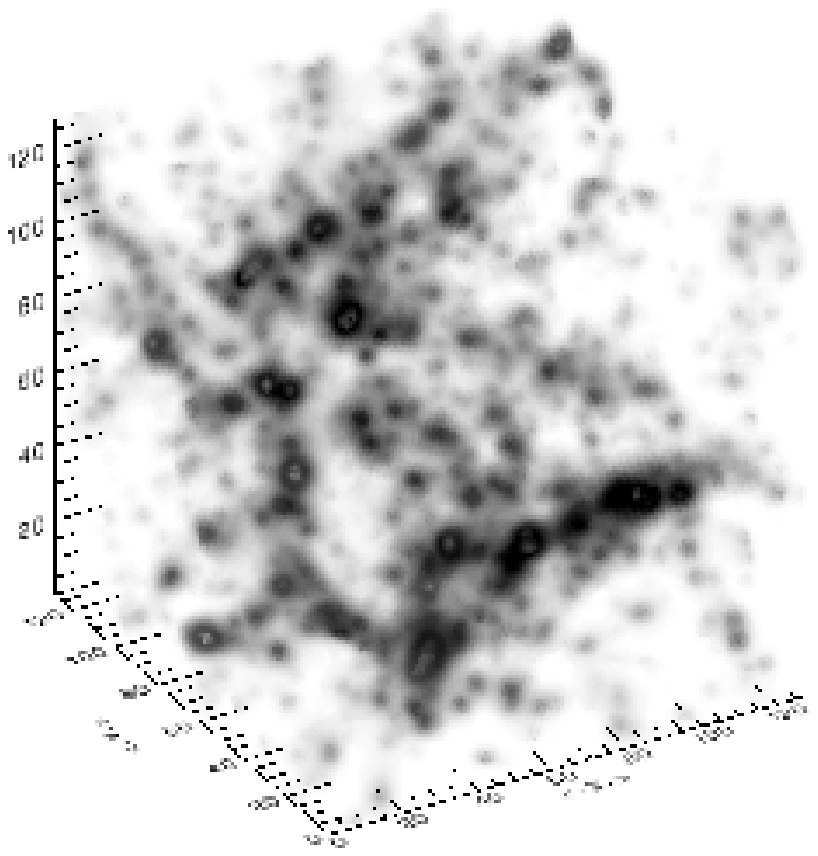}}
}
\caption{Simulated data filtered by a Gaussian filter (two left
panels corresponding respectively to $\sigma=1$ and $\sigma=3$)
and using the wavelet algorithm described in the text (right
panel).} 
\label{fig_lss_wt}
\centerline{ 
\hbox{
\psfig{figure=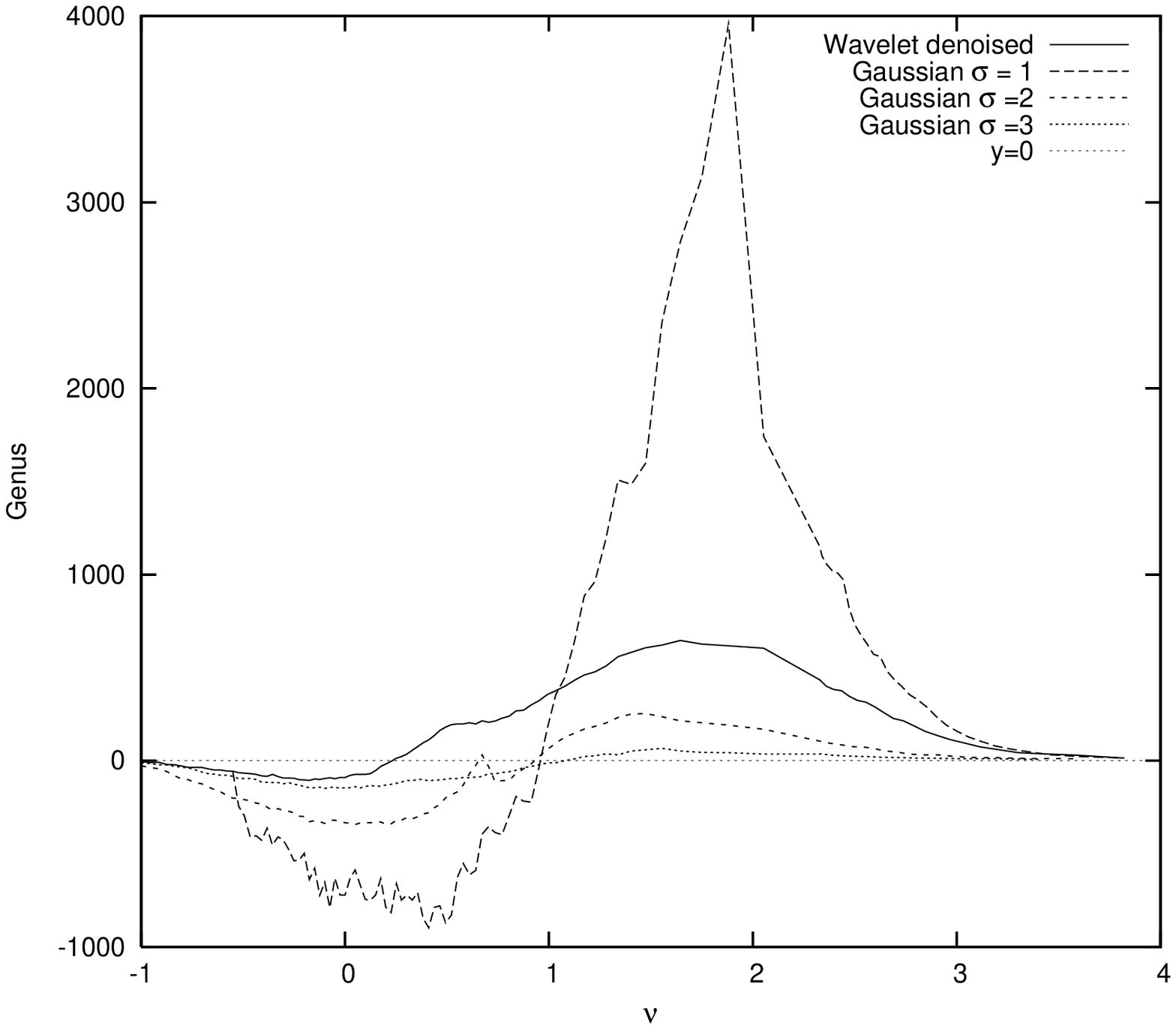,bbllx=1.5cm,bblly=2cm,bburx=19.5cm,bbury=19cm,width=12cm,height=7.5cm,clip=}
}} 
\caption{The genus curve for the N-body model shown in
Fig.~\ref{fig_lss} convolving the data with a Gaussian filter with
different values of $\sigma$ and the genus for the wavelet
filtered data set.} 
\label{genus}
\end{figure}

Fig.~\ref{fig_lss_wt} shows, at the right panel, the same data set
filtered by the 3D wavelet transform, using the algorithm
described previously. The two left panels correspond to Gaussian
smoothing
\begin{equation}
W({\bf x})= \frac{1}{(2\pi)^{3/2} \sigma^3} \exp \left ( -
\frac{{\bf x}^2}{2\sigma^2} \right ).
\end{equation}
with $\sigma=1$ and $\sigma=3$. We can see that when the bandwidth
is too small the discreteness and the noise dominate the density
reconstructed field, while large value of $\sigma$ erase all the
small scale features of the distribution. This is illustrated in
Fig.~\ref{genus}, where we can see the strong dependence of the
genus curve with the width of the Gaussian filter, being its
interpretation completely dependent of the choice of the
bandwidth. The wavelet reconstructed density field keeps
information at all scales due to its multiscale nature. This is
observed in the 3D image at the right panel of
Fig.~\ref{fig_lss_wt}, where we see how large filaments, big
clusters and walls coexist with small scale features such as the
density enhancement around groups and small clusters. The genus
curve of this adaptive reconstructed density field is much more
informative because it does not depend of the particular choice of
the filter radius.

\section{The 3D Ridgelet Transform}
\subsection{The 2D Ridgelet Transform}
The two-dimensional continuous ridgelet transform of
a function $f \in L^2(\bR^2)$ is
defined as follows \cite{cur:candes99_1}.\\
We first select a smooth function $\psi \in L^2(\bR)$, we assume that $\psi$
satisfies the \textit {admissibility} condition

\begin{equation}
\label{eq:admissibility}
\int |\hat{\psi}(\xi)|^2/|\xi| \, d\xi < \infty,
\end{equation}
which holds if $\psi$ has a sufficient decay and a vanishing mean $\int
\psi(t) dt = 0$ ($\psi$ can be normalized so that it has unit energy
$1/(2\pi)\int |\hat{\psi}(\xi)|^2 d\xi = 1$).\\
For each $a > 0$, $b \in \bR$ and $\theta_1 \in [0,2\pi[$, we
define the {\em ridgelet} by

\noindent {\parindent=4cm \indent \(
\begin{array}{ll}
\label{eq:ridgelet}
       \psi_{a,b,\theta_1}: &\bR^2 \goto \bR \\
       &\psi_{a,b,\theta_1} (x_1,x_2) = a^{-1/2} \cdot
       \psi( (x_1 \cos\theta_1 + x_2 \sin\theta_1  - b)/a);
\end{array} \)
}\\
Given a function $f \in L^2(\bR^2)$, we define its ridgelet
coefficients by:

\noindent {\parindent=4cm \indent\(
\begin{array}{lll}
        \cR_f: &\bR^3 \goto \bR \\
               &\cR_f(a,b,\theta_1) = \int \overline{\psi}_{a,b,\theta_1}
           (\mathbf{x}) f(\mathbf{x}) d\mathbf{x}.
\end{array}\)
}

\begin{figure}[htb]
\centerline{
\hbox{
\psfig{figure=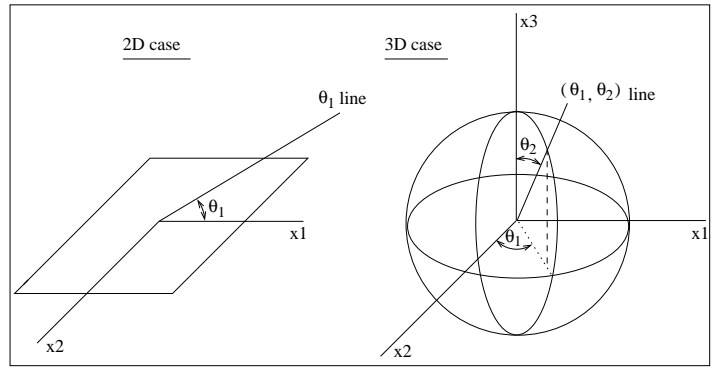}
}}
\caption{Definition of angle1 $\theta_1$ and $\theta_2$ in $\bR^2$ and $\bR^3$ }
\label{fig_angle}
\end{figure}

It has been shown \cite{cur:candes99_1} that the ridgelet transform
is precisely
the application of a 1-dimensional wavelet transform to the slices
of the Radon transform (where the angular variable $\theta_1$ is
constant). This method is therefore optimal to detect lines of the
size of the image (the integration increase as the length of the line).
More details on the implementation of the digital ridgelet transform
can be found in \cite{starck:sta01_3}.

\begin{figure}[htb]
\centerline{
\hbox{
\psfig{figure=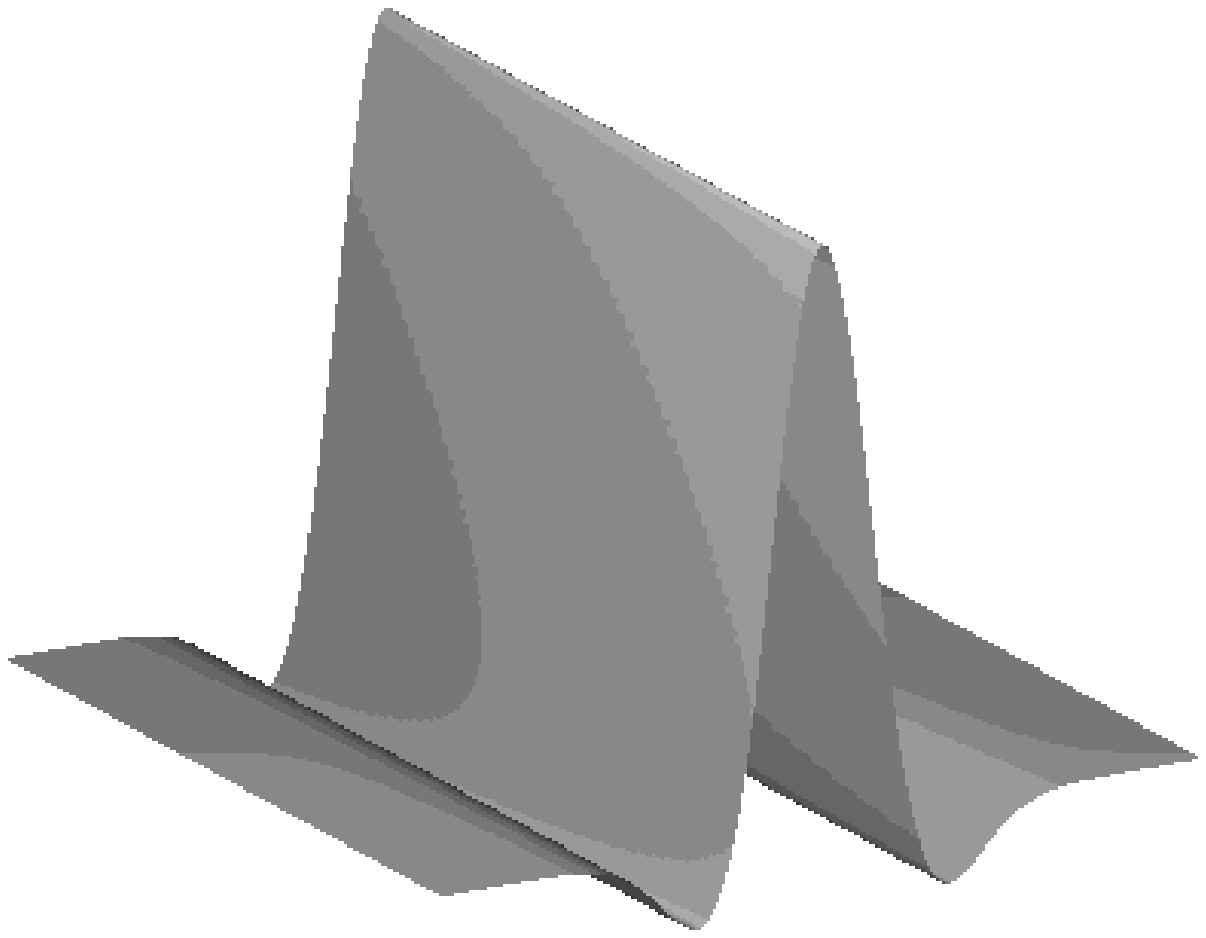,bbllx=4.cm,bblly=13cm,bburx=17cm,bbury=25cm,width=6.cm,height=6.cm,clip=}
\psfig{figure=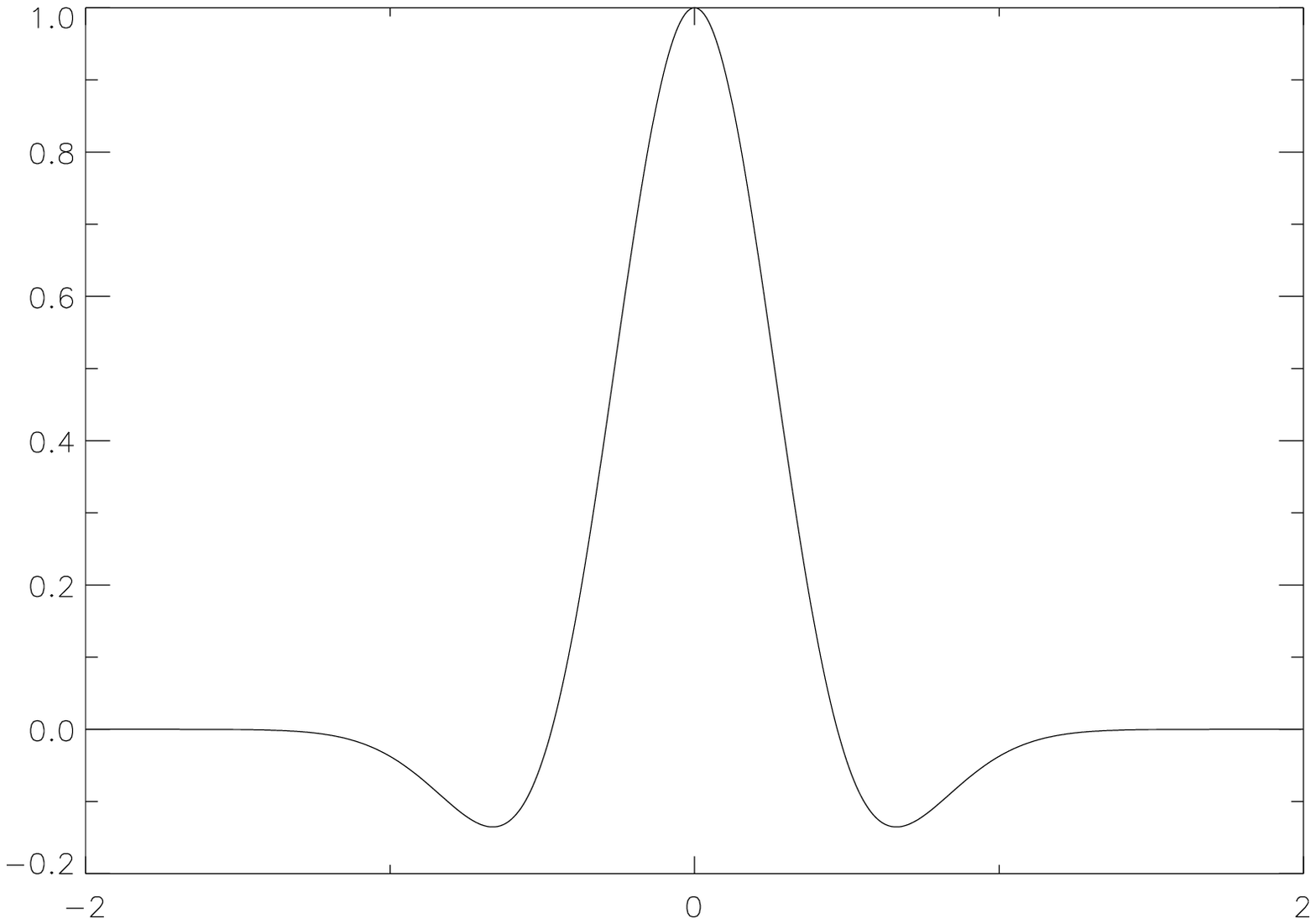,bbllx=3cm,bblly=13cm,bburx=19.5cm,bbury=25cm,width=6.cm,height=6.cm,clip=}
}}
\caption{Example of 2D ridgelet function.}
\label{fig_rid_function}
\end{figure}
Figure~\ref{fig_rid_function} (left)
shows an example ridgelet function.
This function is constant along lines $x_1 \cos\theta + x_2 \sin\theta
= const$. Transverse to these ridges it is a wavelet
(see figure~\ref{fig_rid_function} (right).

\subsection{From 2D to 3D}

The three-dimensional continuous ridgelet transform of
a function $f \in L^2(\bR^3)$ is given by:

\noindent {\parindent=1.5cm \indent\(
\begin{array}{lll}
        \cR_f: &\bR^4 \goto \bR \\
               &\cR_f(a,b,\theta_1,\theta_2) = \int
           \overline{\psi}_{a,b,\theta_1,\theta_2}
           (\mathbf{x}) f(\mathbf{x}) d\mathbf{x}.
\end{array}\)
}\\
where $a > 0$, $b \in \bR$ , $\theta_1 \in [0,2\pi[$ and
$\theta_2 \in [0,\pi[$.\\
The {\em ridgelet} function is defined by:

\noindent {\parindent=1.5cm \indent \(
\begin{array}{ll}
\label{eq:beamlet}
       \psi_{a,b,\theta_1,\theta_2}: &\bR^3 \goto \bR \\
       &\psi_{a,b,\theta_1,\theta_2} (x_1,x_2,x_3) = a^{-1/2} \cdot
       \psi( (  x_1 \cos\theta_1\cos\theta_2 + x_2 \sin\theta_1\cos\theta_2
              + x_3 \sin\theta_2 -b)/a);
\end{array} \)
}

\begin{figure}[htb]
\centerline{
\hbox{
\psfig{figure=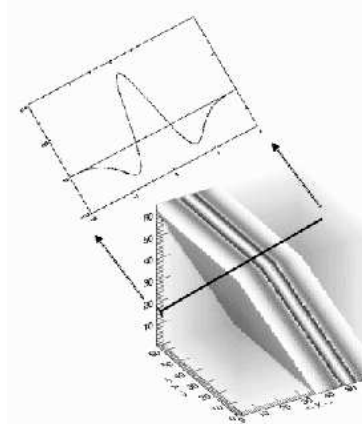,bbllx=1cm,bblly=7cm,bburx=15cm,bbury=23cm,width=7cm,height=8cm,clip=}
}}
\caption{Example of ridgelet function.}
\label{fig_rid3d}
\end{figure}
Figure~\ref{fig_rid3d} shows an example of ridgelet function. It is a
wavelet function in the direction defined by the line $(\theta_1, theta_2)$,
and it is constant along the orthogonal plane to this line.


\begin{figure}[htb]
\centerline{
\hbox{
\psfig{figure=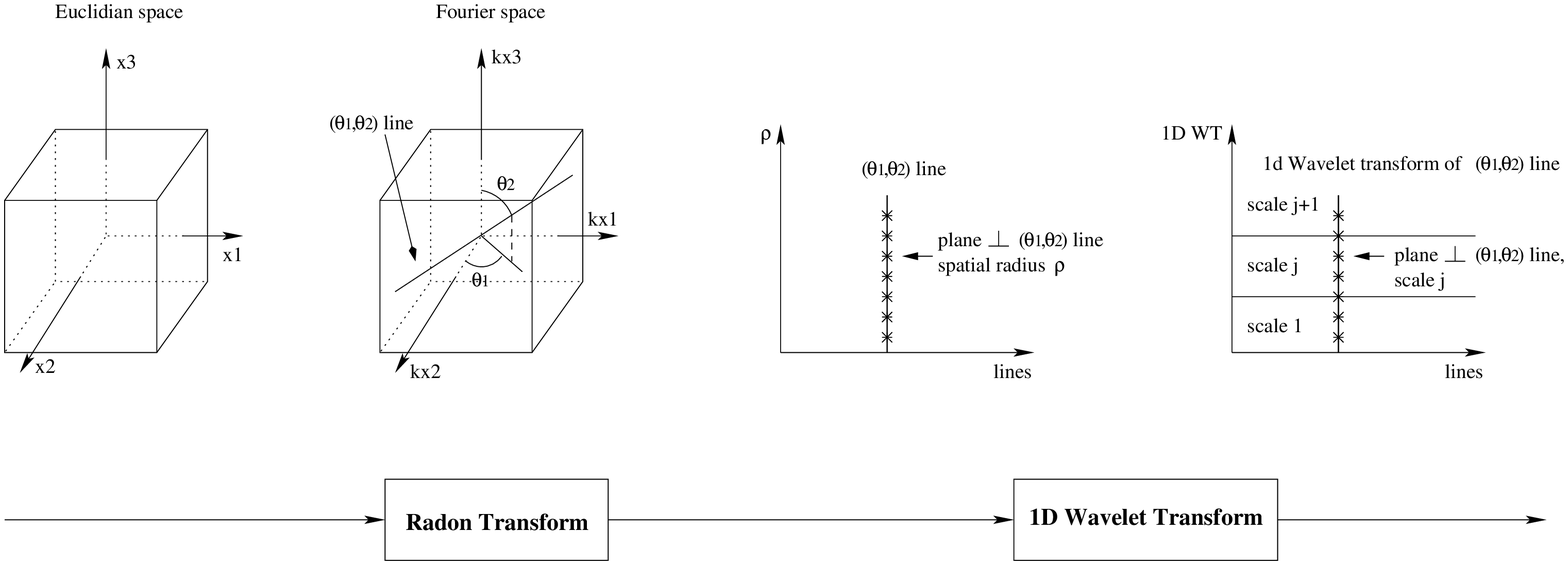,width=15cm,height=6.5cm,clip=}
}}
\caption{3D ridgelet transform flowgraph.}
\label{fig_rid3dtrans}
\end{figure}

As in the 2D case, the 3D ridgelet transform can be built by extracting
lines in the Fourier domain. Let $c(i_1,i_2,i_3)$ a cube of size $(N,N,N)$,
the algorithm consists in the following steps:

\begin{enumerate}
\item {\em 3D-FFT}. Compute $\hat{c}(k_1,k_2,k_3)$, the three-dimensional
      FFT of the cube $c(i_1,i_2,i_3)$.
\item {\em Cartesian to Spherical Conversion}. Using an interpolation
      scheme, substitute the sampled values of $\hat{c}$
      obtained on the Cartesian coordinate system $(k_1,k_2,k_3)$ with sampled values of
      on a spherical coordinate system $(\theta_1,\theta_2,\rho)$.
\item {\em Extract lines}. Extract the $3N^2$ lines (size $N$) passing through
      the origin and the boundary of $\hat{c}$.
\item {\em 1D-IFFT}. Compute the one-dimensional inverse FFT on each line.
\item {\em 1D-WT}. Compute the one-dimensional wavelet transform on each line.
\end{enumerate}
Figure~\ref{fig_rid3dtrans} the 3D ridgelet transform flowgraph.
The 3D ridgelet transform allows us to detect sheets in a cube.

\subsubsection*{Local 3D Ridgelet Transform}

The ridgelet transform is optimal to find sheets of the size of
the cube. To detect smaller sheets, a partitioning must be
introduced \cite{cur:candes99_4}. The cube $c$ is decomposed into
blocks of lower side-length $b$ so that for a $N*N*N$ cube, we
count $N/b$ blocks in each direction. After the block
partitioning, The detection is therefore optimal for sheets of
size $b \times b$ and of thickness  $a_j$, $a_j$ corresponding to
the different dyadic scales used in the transformation.

\section{The 3D Beamlet Transform}
\subsection{Definition}
The X-ray transform of a continuum function $f(x,y,z)$ with
$(x,y,z) \in \bR^3$ is defined by
\begin{eqnarray}
 (Xf)(L)  = \int_L f(p) dp
\end{eqnarray}
where $L$ is a line in $\bR^3$, and $p$ is a variable indexing
points in the line. The transformation contains all line integrals
of $f$. The Beamlet Transform (BT)  can be seen as a multiscale
digital X-ray transform. It is multiscale transform because, in
addition to the multiorientation and multilocation line integral
calculation, it integrated also over line segments at different
length. The 3D BT is an extension to the 2D BT, proposed by Donoho
and Huo \cite{cur:donoho_01}.

\subsubsection*{The system of 3D beams}

The first choice to consider is the line segment set. We would
like to have an expressive set of line segments in the sense that
it includes line segments with various lengths, locations and
orientations lying inside a 3D volume and at the same time has
reasonable size.

A seemingly natural candidate for the set of line segments is the
family of all line segments between any voxel corner and any other
voxel corner, the set of {\it 3-D beams}. The beams set is
expressive but can be of huge cardinality for even moderate
resolutions. For a 3D data set with $n^3$ voxels we get $O(n^6)$
3D beams - So that is clearly infeasible to use the collection of
3-D beams as a basic data structure since any algorithm based on
this set will have a complexity with lower bound of $n^6$ and
hence unworkable for typical sizes 3-D images.

\subsection{The Beamlet System}

A dyadic cube $C(k_1,k_2,k_3,j) \subset [0,1]^3$ is  the collection of points
\[
\{(x_1,x_2,x_3):[k_1/2^j,(k_1+1)/2^j]\times [k_2/2^j,(k_2+1)/2^j]
\times [k_3/2^j,(k_3+1)/2^j]\}
\]
where $0\le k_1,k_2,k_3 < 2^j$ for an integer $j \ge 0$.
We will refer to $j$ as the scale of the dyadic cube

Such cubes can be viewed as descended from the
unit cube $C(0,0,0,0)= [0,1]^3$ by recursive partitioning.
Hence, the result of splitting $C(0,0,0,0)$ in
half along each axis is the eight cubes
$C(k_1,k_2,k_3,1)$ where $k_i \in \{0,1\}$,
splitting those in half along each axis we get the $64$ subcubes
$C(k_1,k_2,k_3,2)$ where $k_i \in \{0,1,2,3\}$,
and if we decompose the unit cube into
$n^3$ voxels using a uniform $n$-by-$n$-by-$n$ grid
with $n=2^J$ dyadic, then the individual voxels are
the $n^3$ cells $C(k_1,k_2,k_3,J)$,
$0\le k_1,k_2,k_3 < n.$

\begin{figure}[htb]
\centering
\includegraphics[height=2in]{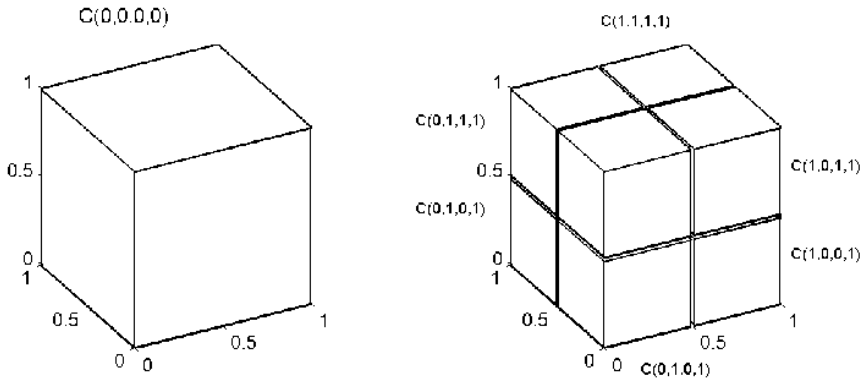}
\caption []{Dyadic cubes}
\label{fig:dcubes}
\end{figure}

Associated to each dyadic cube we can build a system of line
segments that have both of their end-points lying on the cube
boundary. We call each such segment {\it A beamlet}. If we
consider all pairs of boundary voxel corners we get $O(n^4)$
beamlets for a dyadic cube with a side length of $n$ voxels, we
will work with a slightly different system in which each line is
associated with a slope and an intercept instead of its end-points
as will be explained below. However, we will still have $O(n^4)$
cardinality. Assuming a voxel size of $1/n$ we get $J+1$ scales of
dyadic cubes where $n = 2^J$, for any scale $0\le j \le J$ there
are $2^{3j}$ dyadic cubes of scale $j$ and since each dyadic cube
at scale $j$ has a side length of $2^{J-j}$ voxels we get
$O(2^{4(J-j)})$ beamlets associated with the dyadic cube and a
total of $O(2^{4J-j})=O(n^4/2^j)$ beamlets at scale j. If we sum
the number of beamlets at all scales we get $O(n^4)$ beamlets.

\begin{figure}[htb]
\centering
\includegraphics[height=2in]{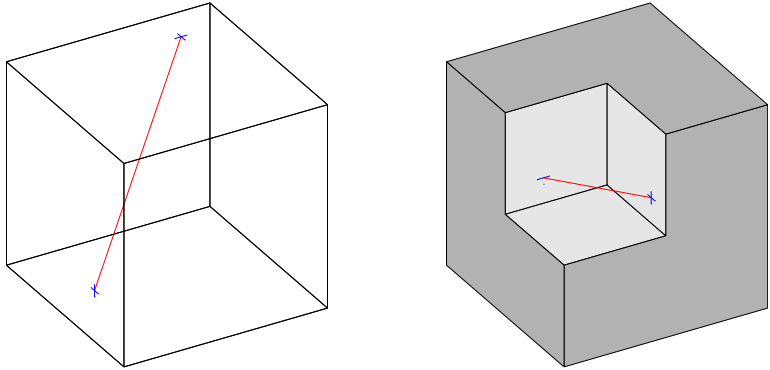}
\caption []{Examples of Beamlets at two different scales.
(a) Scale 0 (coarsest scale) (b) Scale 1 (next finer scale).}
\label{fig:beamlets}
\end{figure}

We have constructed above a multi-scale arrangement of line
segments in 3D with controlled cardinality of $O(n^4)$, the scale
of a beamlet is defined as the scale of the dyadic cube it belongs
to so lower scales correspond to longer line segments and finer
scales correspond to shorter line segments. Figure
\ref{fig:beamlets} shows 2 beamlets at different scales.

To construct the set of beamlets for a given dyadic cube we use
the slope-intercept pairs system. For a data cube of $n \times n
\times n$ voxels consider a coordinate system with the cube center
of mass at the origin and a unit length for a voxel. Hence, for
$(x,y,z)$ in the data cube we have $|x|, |y|, |z| \leq n/2$. We
can consider three kinds of lines: {\it $x$-driven, $y$-driven,
and $z$-driven}, depending on which axis provides the shallowest
slopes. An $x$-driven line takes the form
\[
       z = s_z x + t_z, \quad y = s_y x + t_y
\]
with slopes $s_z$,$s_y$, and intercepts $t_z$
and $t_y$. Here the slopes $|s_z|, |s_y| \leq 1$.
$y$- and $z$-driven lines are defined with
an interchange of roles between $x$ and $y$
or $z$, as the case may be.

We will consider the family of lines generated
by this, where the slopes and intercepts
run through an equispaced family:
\[
      s_x, s_y, s_z \in \{ 2\ell/n :\ell = -n/2 , \dots , n/2-1 \}, \qquad
       t_x, t_y, t_z \in \{ \ell: -n/2,\dots,n/2-1 \} .
\]


Using the above family of lines for a given data cube we can
define the set of beamlets belong to the cube to be the set of
line segment obtained by taking the intersection of each line with
the cube.

With the choice of indices range above we get that all beamlets
associated with a data cube of size $n$ have lengths bigger than
$n/2$, half of the cube length and $\sqrt{3} n$, the cube main
diagonal length.

\subsubsection*{Computational aspects}

The Beamlet coefficients are the line integrals over the set of
beamlets. A digital 3-D image can be regarded as a 3-D piece-wise
constant function and each line integral is just a weighted sum of
the voxel intensities along the corresponding line segment. Donoho
and Levi \cite{cur:donoho02} discuss in detail different
approaches for computing line integrals in a 3-D digital image.
Computing the beamlet coefficients for real applications data sets
can be a challenging computational task since for a data cube with
$n\times n\times n$ voxels we have to compute $O(n^4)$
coefficients. By developing efficient cache aware algorithms we
are able to handle 3-D data sets of size up to $n=256$ on a single
fast machine in less than a day running time. We will mention that
in many cases there is no interest in the coarsest scales
coefficient that consumes most of the computation time and in that
case the over all running time can be significantly faster. The
algorithms can also be easily implemented on a parallel machine of
a computer cluster using a system such as MPI in order to solve
bigger problems.

\subsection{The FFT-based transformation}

Let $\psi \in L^2(\bR^2)$ a smooth function satisfying the
\textit {admissibility} condition (a 2D wavelet function),
the three-dimensional continuous
beamlet transform of  a function $f \in L^2(\bR^3)$is given by:

\noindent {\parindent=1.5cm \indent\(
\begin{array}{lll}
        \mathcal{B}_f: &\bR^5 \goto \bR \\
               &\mathcal{B}_f(a,b_1,b_2,\theta_1,\theta_2) = \int
           \overline{\psi}_{a,b,\theta_1,\theta_2}
           (\mathbf{x}) f(\mathbf{x}) d\mathbf{x}.
\end{array}\)
}\\
where $a>0$, $b_1,b_2 \in \bR$ ,$\theta_1 \in [0,2\pi[$ and
$\theta_2 \in [0,\pi[$.\\
The {\em beamlet} function is defined by:

\noindent {\parindent=0cm \indent \(
\begin{array}{lll}
\label{eq:beamlet_3d}
       \psi_{a,b_1,b_2,\theta_1,\theta_2}: &\bR^3 \goto \bR \\
       &\psi_{a,b_1,b_2,\theta_1,\theta_2} (x_1,x_2,x_3) = a^{-1/2} \cdot
       \psi( &(-x_1 \sin\theta_1 + x_2 \cos\theta_1 + b_1  )/a, \\
       &     &(  x_1 \cos\theta_1 \cos\theta_2 + x_2 \sin\theta_1 \cos\theta_2
               - x_3 \sin\theta_2 + b_2)/a);
\end{array} \)
}

\begin{figure}[htb]
\centerline{
\hbox{
\psfig{figure=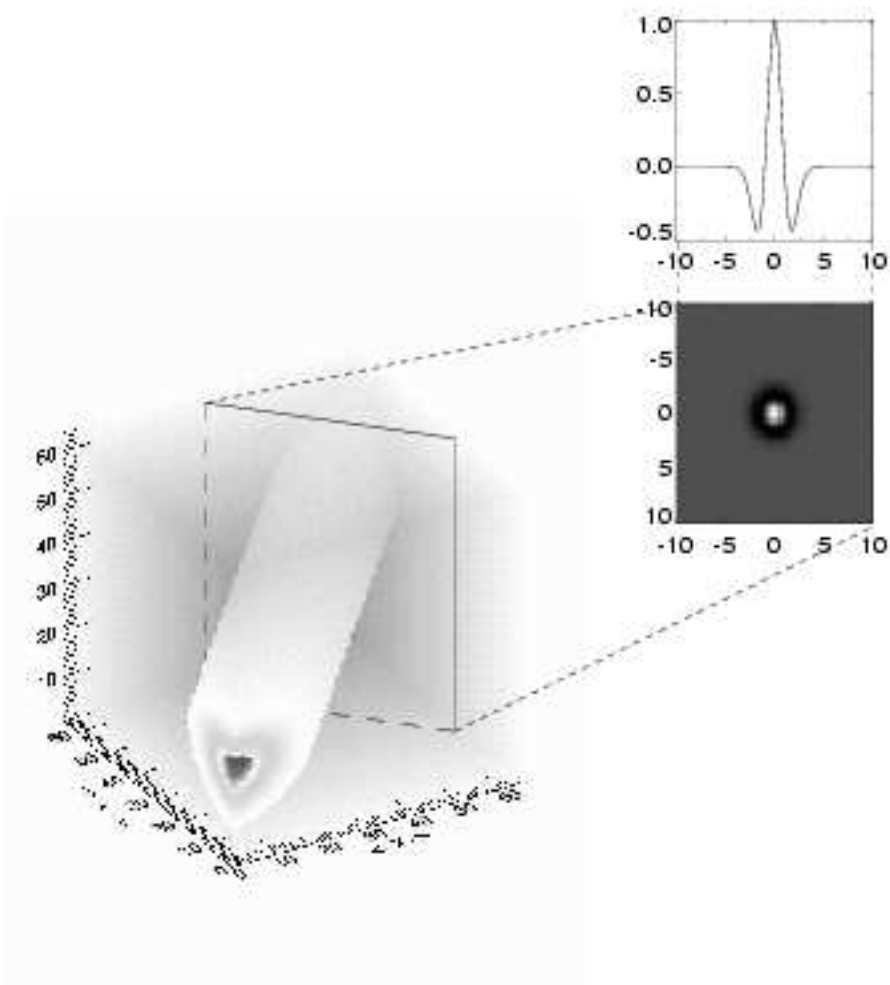}
}}
\caption{Example of beamlet function.}
\label{fig_beam3d}
\end{figure}

Figure~\ref{fig_rid3d} shows an example of beamlet function. It is constant
along lines of direction ($\theta_1$, $\theta_2$), and a 2D wavelet function
along plane orthogonal to this direction.

\begin{figure}[htb]
\centerline{
\hbox{
\psfig{figure=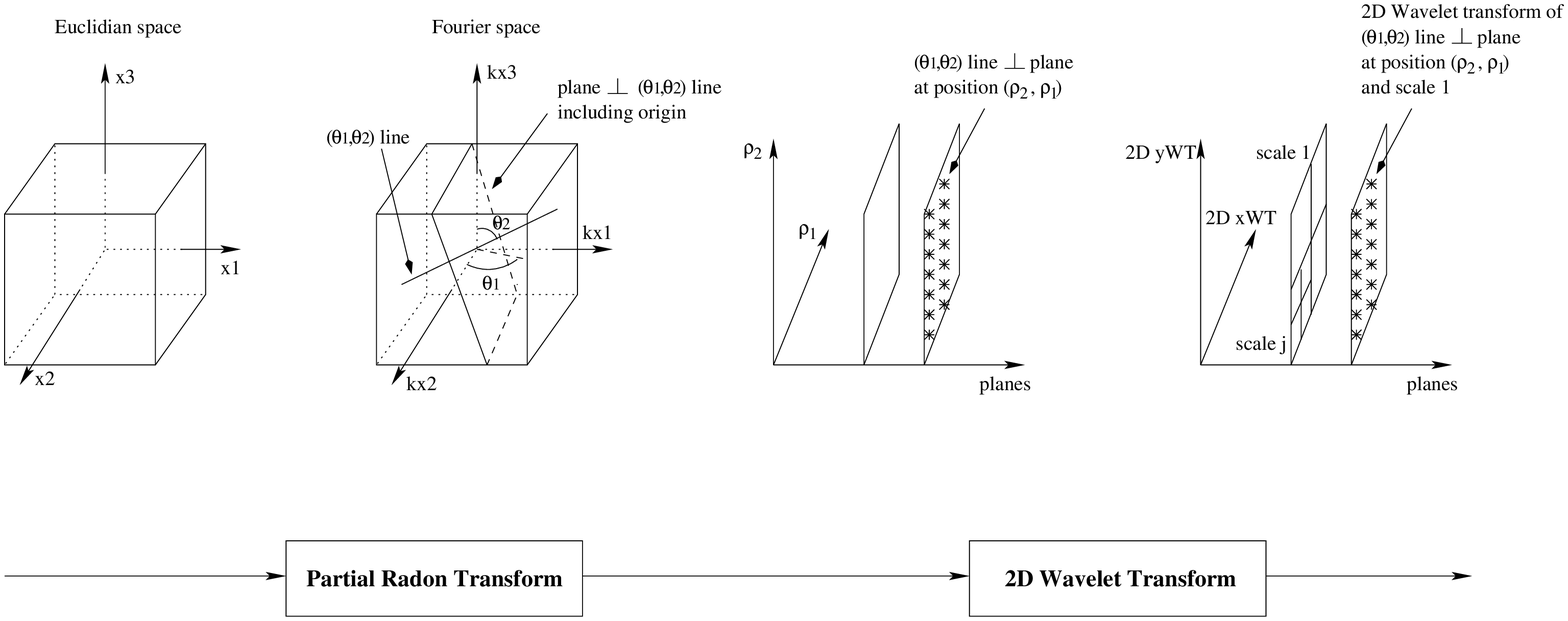,width=15cm,height=6.5cm,clip=}
}}
\caption{3D beamlet transform flowgraph.}
\label{fig_beam3dtrans}
\end{figure}

The 3D beamlet transform can be built using the "Generalized
projection-slice theorem" \cite{ima:zhi00}. Let

\begin{itemize}
\item $f(\mathbf{x})$ an $n$ dimensional function,
\item $\mathcal{R}ad_mf$ its m-dimensional partial radon transform along the first m
      cardinal directions, $m < n$, $\mathcal{R}ad_mf$ is a function of
      $(p,\mu_m;x_{m+1},...,x_n)$, $\boldsymbol {\mu}_m$ a unit
      directional vector in $\mathcal{R}ad_m$ (note that for a given projection angle,
      the $m$ dimensional partial radon transform of $f(\mathbf{x})$ has
      $(n-m)$ untransformated spatial dimension and a (n-m+1) dimensional
      projection profile),
\item $\{\mathcal{F}f\}(\mathbf{k})$ its Fourier transform ($\mathbf{x}$ and
      $\mathbf{k}$ are conjugate variable pairs of $\mathcal{F}$),
\end{itemize}

The Fourier transform of the $m$ dimensional partial radon transform $\cR_mf$
is related to the Fourier transform of $f$ $( \mathcal{F}f )$ by the following relation
\begin{equation}
\label{eq:partradtrans}
\{\mathcal{F}_{n-m+1}\mathcal{R}ad_mf\}(k,k_{m+1},...,k_n) =
\{\mathcal{F}f\}(k\boldsymbol{\mu}_m,k_{m+1},...,k_n)
\end{equation}

Let $c(i_1,i_2,i_3)$ a cube of size $(N,N,N)$, the Beamlet algorithm consists
in the following steps:
\begin{enumerate}
\item {\em 3D-FFT}. Compute $\hat{c}(k_1,k_2,k_3)$, the three-dimensional
      FFT of the cube $c(i_1,i_2,i_3)$.
\item {\em Cartesian to Spherical Conversion}. Using an interpolation
      scheme, substitute the sampled values of $\hat{c}$
      obtained on the Cartesian coordinate system $(k_1,k_2,k_3)$
      with sampled values of
      on a spherical coordinate system $(\theta_1,\theta_2,\rho)$.
\item {\em Extract planes}. Extract the $3N^2$ planes (of size $N \times N$) passing through
      the origin (each line used in the 3D ridgelet transform defines set of
      orthogonal planes, we take the one which include the origin).
\item {\em 2D-IFFT}. Compute the two-dimensional inverse FFT on each plane.
\item {\em 2D-WT}. Compute the two-dimensional wavelet transform on each plane.
\end{enumerate}
Figure~\ref{fig_beam3dtrans} the 3D beamlet transform flowgraph.
The 3D beamlet transform allows us to detect filament in a cube.
The beamlet transform  algorithm presented in this section differs from
the one presented in \cite{starck:spie02a}, and relation between both
algorithms is given in \cite{cur:donoho02}.

\section{Experiments}
\subsection{Experiment 1}
We have simulated three data set containing respectively a cluster, a plane
and a line. On each data set, Poisson noise have been added with eight
different background levels. Then we have applied the three transforms on the
24 simulated data set. The coefficients distribution related to each
transformation is normalized using twenty realizations of a 3D flat distribution
with a Poisson noise and which have the same number of counts as in the data.

\begin{figure}[ht]
\centering
\resizebox{!}{0.2\textheight}{\includegraphics*{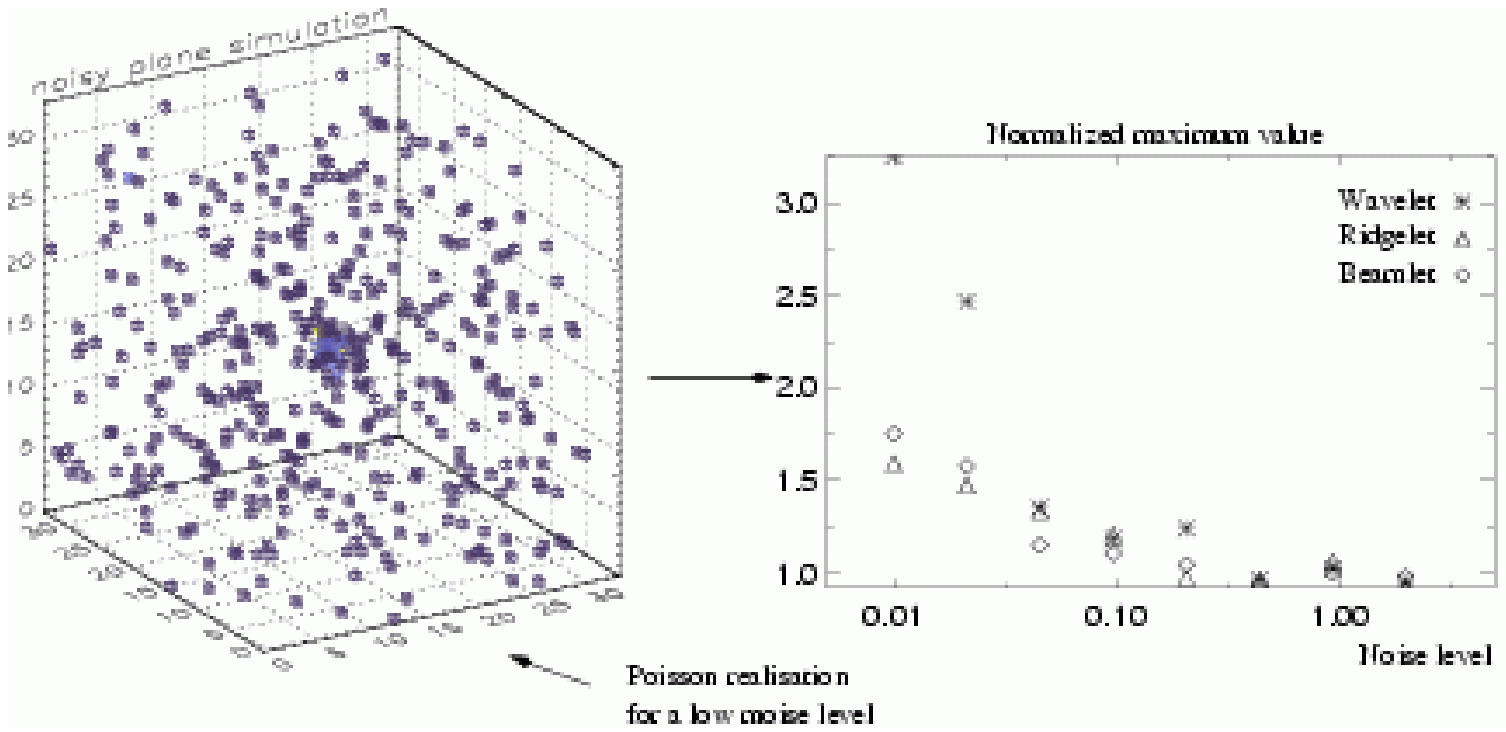}}\\
\resizebox{!}{0.2\textheight}{\includegraphics*{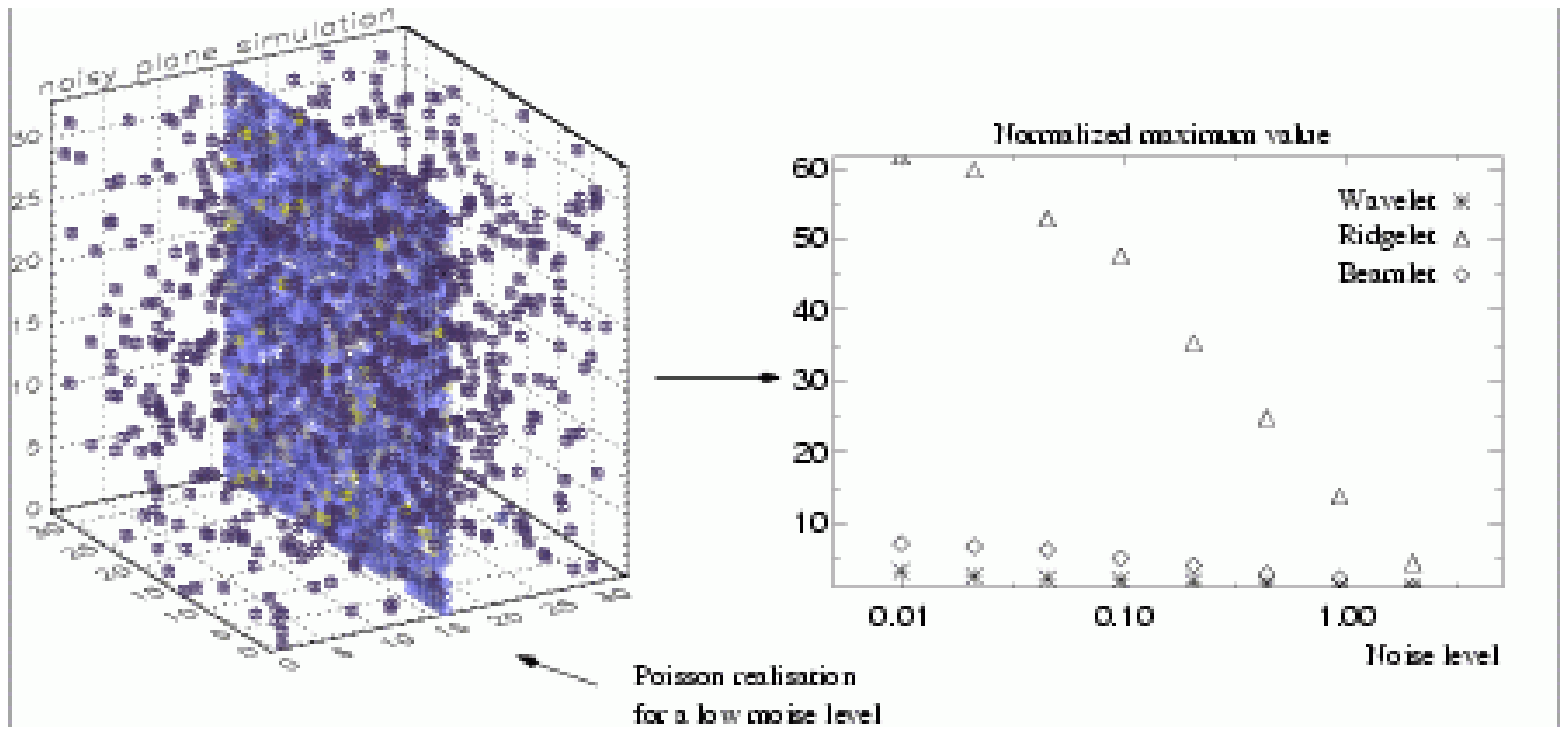}}\\
\resizebox{!}{0.2\textheight}{\includegraphics*{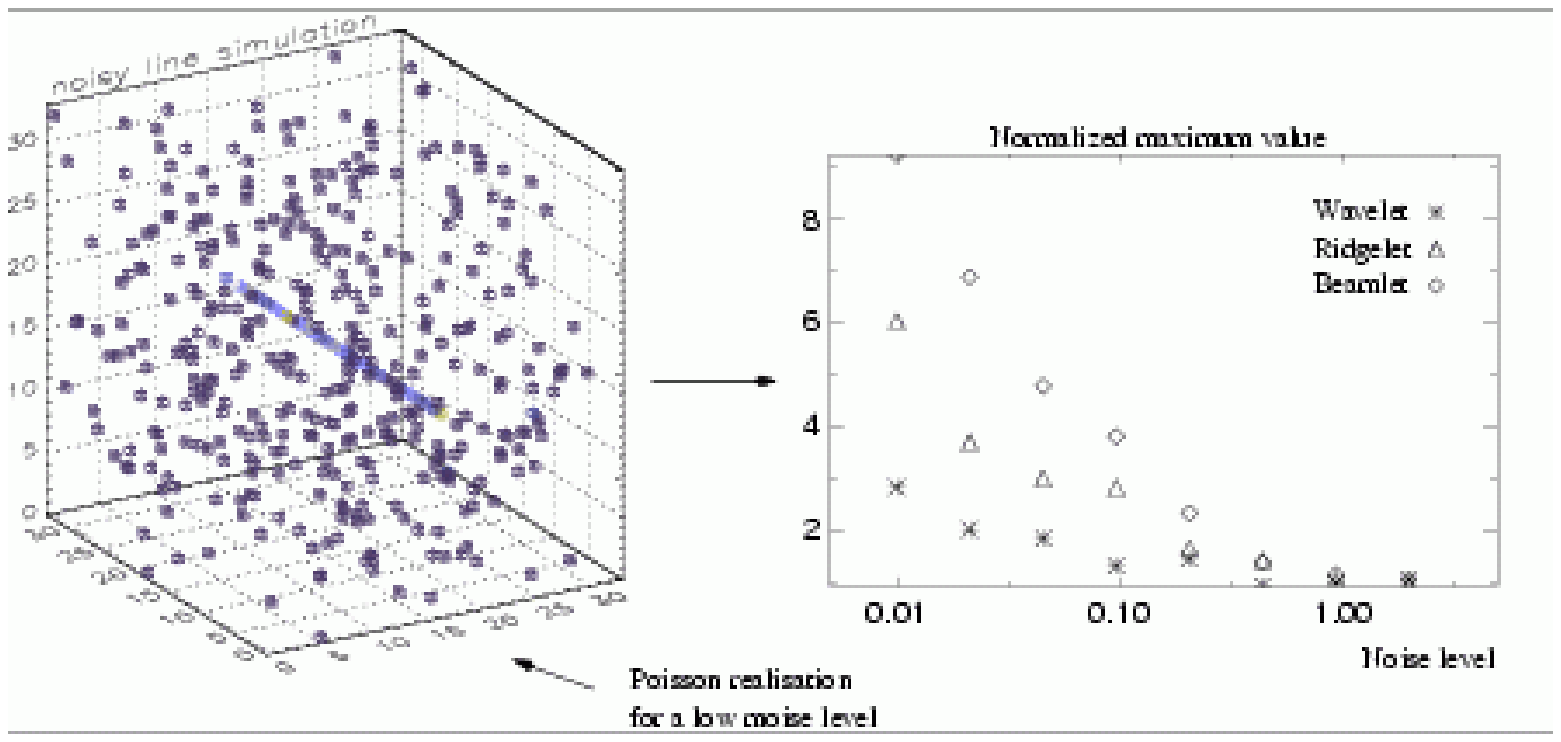}}\\
\caption{Simulation of cubes containing a cluster (top), a plane (middle) and
a line (bottom).}
\label{fig_expalltrans}
\end{figure}

Figure~\ref{fig_expalltrans} shows, from top to bottom, the maximum value of the
normalized distribution versus the noise level for our three simulated data
set. As expected, wavelets, ridgelets and beamlets are respectively the
best for clusters, sheets and lines detection.
It must also be underlined that a feature can be detected with a very
high signal-to-noise ratio in a given basis, and and not detected in another basis.
For example, the wall is detected at more than $60\sigma$ by the ridgelet transform,
and less than $5\sigma$ by the wavelet transform. The line is detected almost at
$10\sigma$ by the beamlet transform, and is under a  $3\sigma$ detection level
using wavelets. These results shows the importance of using several transforms for
an optimal detection of all features contained in a data set.

\subsection{Experiment 2}

\begin{figure}[htb]
\centering
\resizebox{.35\textwidth}{.4\textwidth}{\includegraphics*{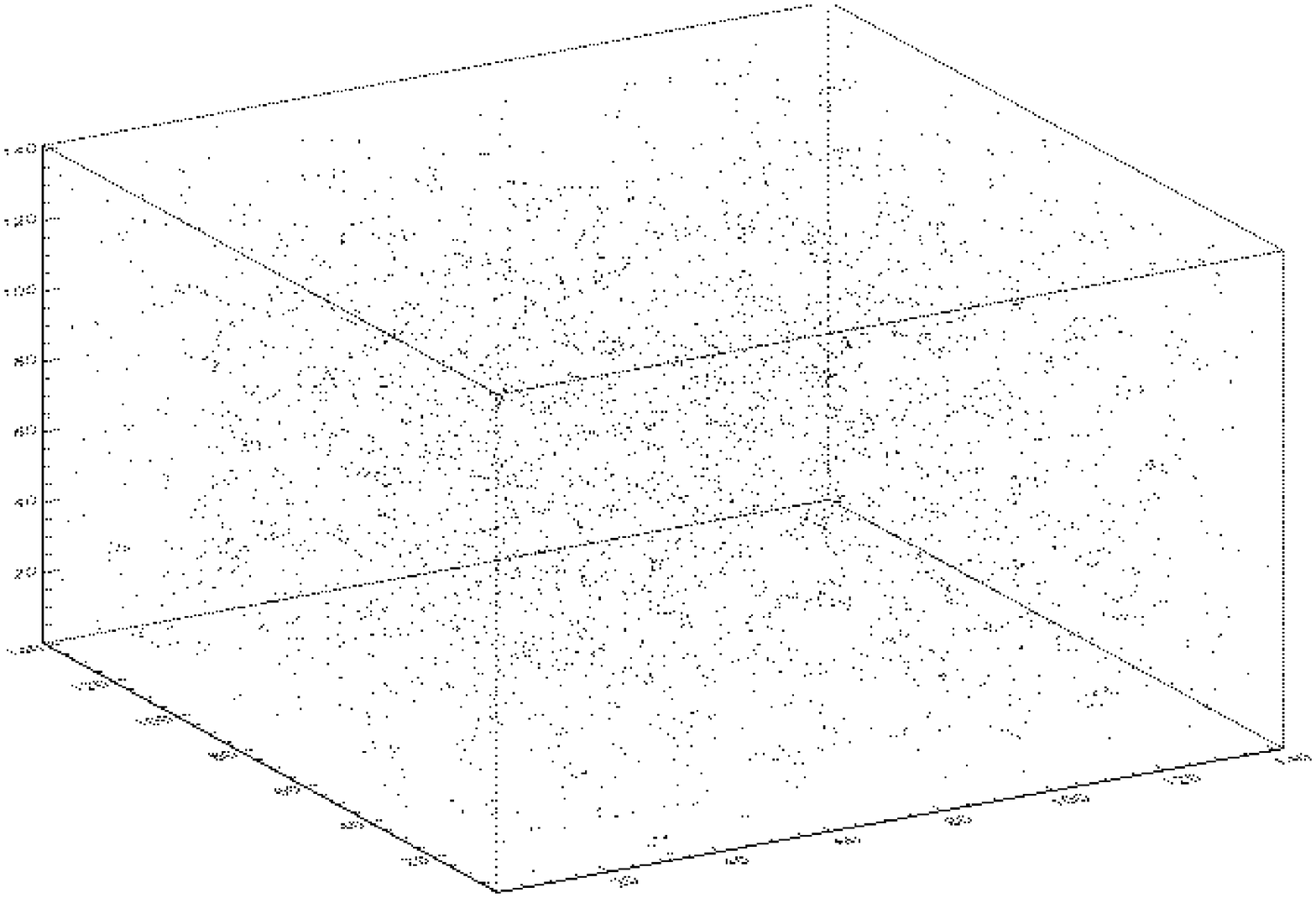}}
\resizebox{.35\textwidth}{.4\textwidth}{\includegraphics*{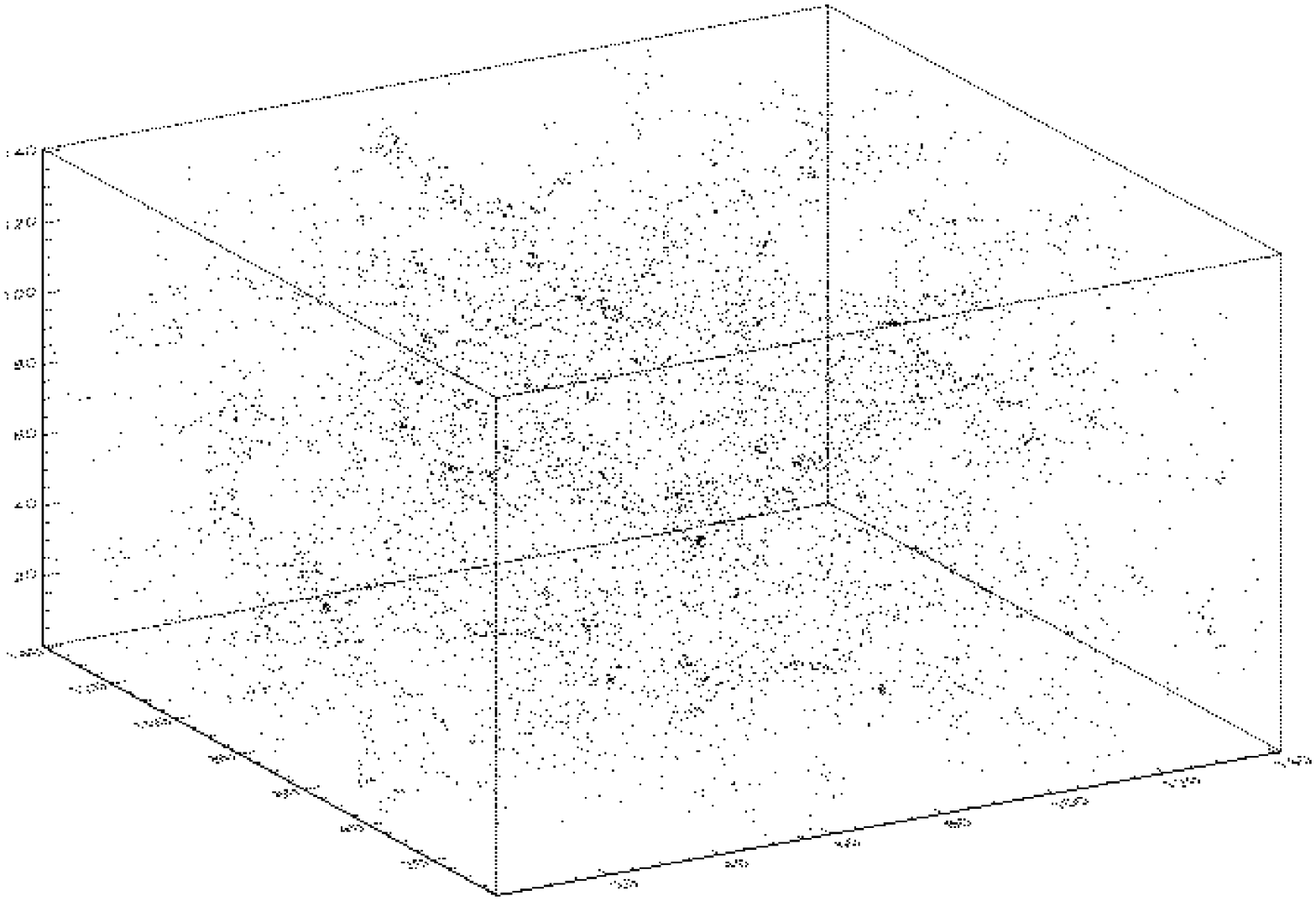}}\\
\vspace*{10pt}

\centerline{\hbox{%
\psfig{figure=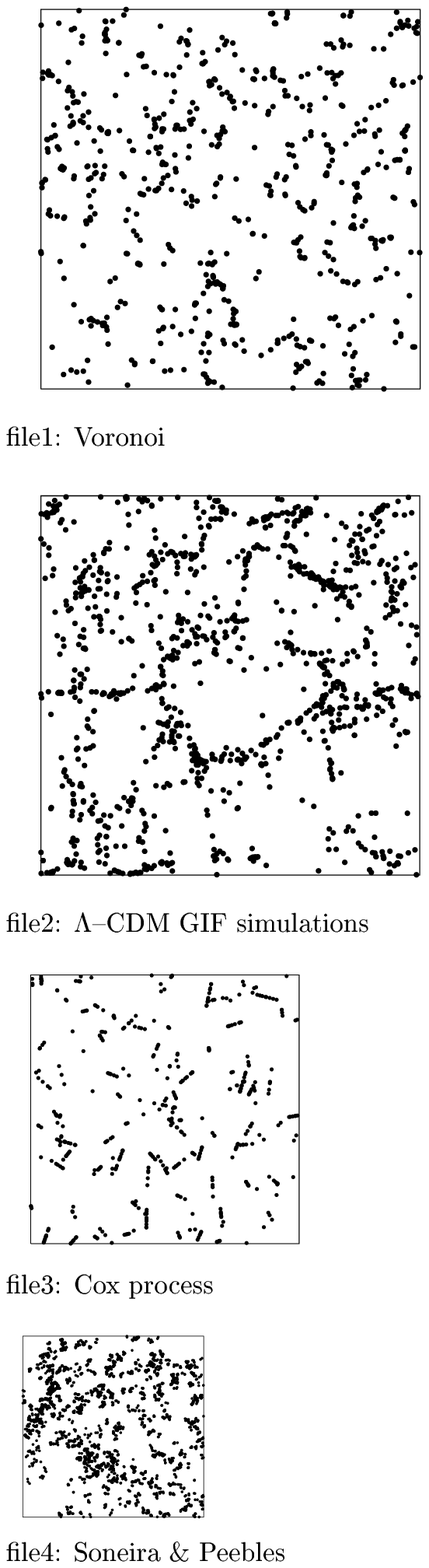,bbllx=2.1cm,bblly=19.8cm,bburx=7.3cm,bbury=25cm,width=.3\textwidth,clip=}
\psfig{figure=fig_slice_voronoi_lcdm.ps,bbllx=2.1cm,bblly=13.5cm,bburx=7.3cm,bbury=18.7cm,width=.3\textwidth,clip=}
}}
\caption{Simulated data sets. Top, the Voronoi vertices point
pattern (left) and the galaxies of the GIF $\Lambda$-CDM N-body
simulation (right). The bottom panels show one 10 $h^{-1}$ width
slice of the each data set.} \label{fix_sim_process4}
\end{figure}

\begin{figure}[htb]
\centerline{ \hbox{
\psfig{figure=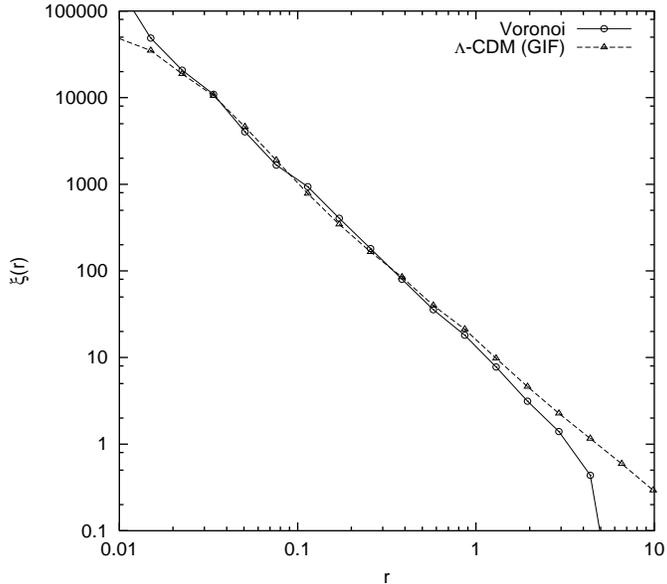,bbllx=1.5cm,bblly=1.5cm,bburx=19cm,bbury=17.5cm,width=9cm,height=8cm,clip=}
}} \caption{The two-point correlation function of the Voronoi
vertices process and the GIF $\Lambda$-CDM N-body simulation. They
are very similar in the range [0.02,2] $h^{-1}$ Mpc.}
\label{fix_sim_cox}
\end{figure}

\begin{figure}[htb]
\centerline{
\hbox{
\psfig{figure=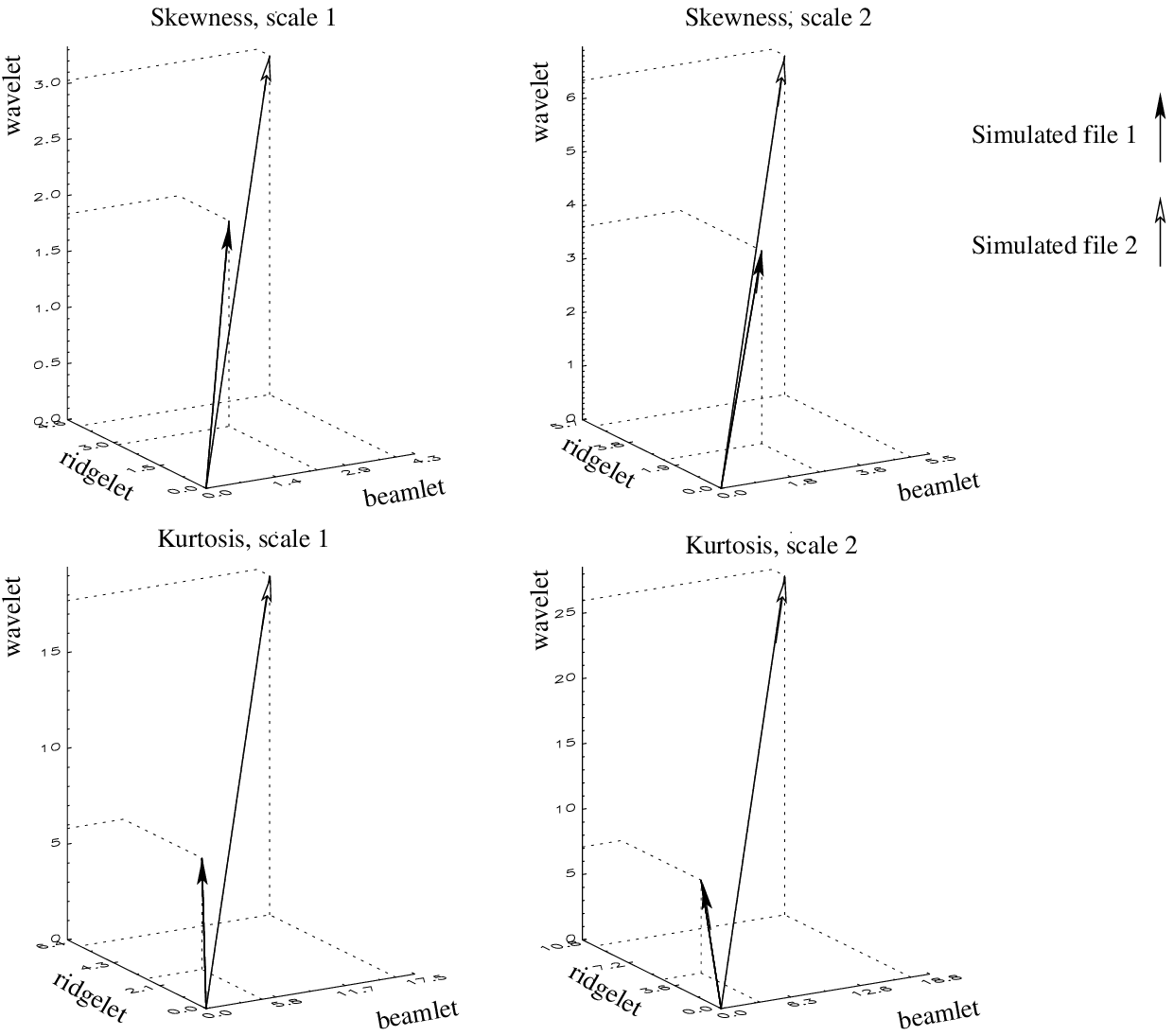}
}}
\caption{Skewness and kurtosis for the two simulated data set.}
\label{Skew_Kurt_1}
\end{figure}

We use here two simulated data sets to illustrate the
discriminative power of the multiscale methods. The first one is a
simulation from stochastic geometry. It is based on a Voronoi
model. The second one is a mock catalog of the galaxy distribution
drawn from a $\Lambda$-CDM N-body cosmological model\cite{gif}.
Both processes have very similar two-point correlation functions
at small scales, although they look quite different and have been
generated following completely different algorithms.
\begin{itemize}
\item the first comes from a Voronoi simulation: We locate a point
in each of the vertices of a Voronoi tessellation of $1.500$ cells
defined by $1500$ nuclei distributed following a binomial process.
There are 10085 vertices lying within a box of 141.4 $h^{-1}$ Mpc
side. \item the second point pattern represents the galaxy
positions extracted from a cosmological $\Lambda$-CDM N-body
simulation. The simulation has been carried out by the Virgo
consortium and related groups (see {\tt
http://www.mpa-garching.mpg.de/Virgo}). The simulation is a
low-density ($\Omega =0.3$) model with cosmological constant
$\Lambda=0.7$. It is, therefore, the closer set to the real galaxy
distribution\cite{gif}. There are 15445 galaxies within a box with
side 141.3 $h^{-1}$ Mpc. Galaxies in this catalog have stellar
masses exceeding $2 \times 10^{10}$ $M_\odot$.
\end{itemize}

Figure~\ref{fix_sim_process4} shows the two simulated data set,
and Figure~\ref{fix_sim_cox} shows the two-point correlation
function curve for the two point processes. The two point fields
are different, but as it can be seen in Fig.~\ref{fix_sim_cox},
both have very similar two-point correlation functions in a huge
range of scales (2 decades).

We have applied the three transforms to each data set,
and we have calculated the skewness vector $S = (s^j_w,s^j_r,s^j_b)$
and the kurtosis vector $K = (k^j_w,k^j_r,k^j_b)$ at each scale $j$.
 $s^j_w,s^j_r,s^j_b$ are respectively the skewness at scale $j$
of the wavelet coefficients, the ridgelet coefficients and the beamlet
coefficients. $k^j_w,k^j_r,k^j_b$ are respectively the kurtosis at scale $j$
of the wavelet coefficients, the ridgelet coefficients and the beamlet
coefficients. Figure~\ref{Skew_Kurt_1} shows the kurtosis and the
skewness vectors of the two data set at the two first scales.
On the contrary to the two-point correlation function, this analysis
shows strong differences between the two data set, particularly
on the wavelet axis, which indicates that the
second data contains more, or with a higher density, clusters
than the first one.

\subsection{Experiment 3}
In this experiment, we have used  a $\Lambda$-CDM simulation based
on a N-body and hydrodynamical code, called RAMSES \cite{astro:teyssier02}.
The code is based on Adaptive Mesh Refinement (AMR) technique,
with a tree-based data structure allowing recursive grid
refinements on a cell-by-cell basis. The simulated data have been
obtained using  $256^3$ particles and $4.1 \times 10^7$ cells
in the AMR grid, reaching a formal resolution of $8192^3$.
The box size was set to $100 h^-1$ Mpc, with the following cosmological
parameters:
\begin{eqnarray}
\Omega_m = 0 .3 & \Omega_{\lambda} = 0.7 & \Omega_{b} = 0.039 \nonumber \\
h = 0.7 & \sigma_8 = 0.92 &
\end{eqnarray}

\begin{figure}[htb]
\centering
\resizebox{!}{0.7\textheight}{\includegraphics*{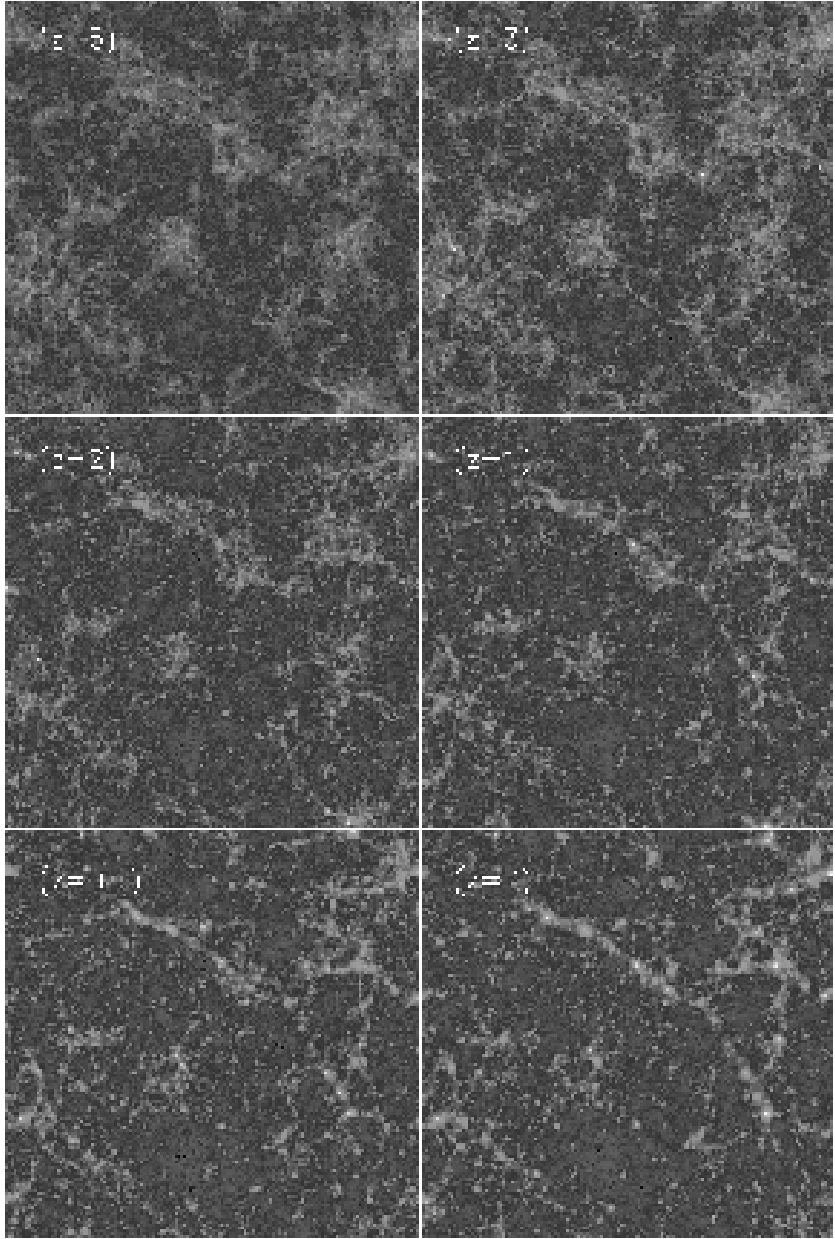}}\\
\caption{$\Lambda$-CDM simulation at different redshifts.}
\label{fig_sim_romain}
\end{figure}
We used the results of this simulation at six different redshifts
($z=5,3,2,1,0.5,0$).
Fig.~\ref{fig_sim_romain} shows a projection of the simulated cubes
along one axis.
We have applied the 3D wavelet transform, the 3D beamlet transform and
the 3D ridgelet transform on the six data set, and we calculate for
each transform the standard deviation of the different scales. We will
note  $\sigma^2_{W,z,j},\sigma^2_{R,z,j},\sigma^2_{B,z,j}$ the
variance of the scale $j$ relative to the transformation
of the cube at redshift $z$ by respectively the wavelet, the ridgelet
and the beamlet transform.

\begin{figure}[htb]
\vbox{
\centerline{
\psfig{figure=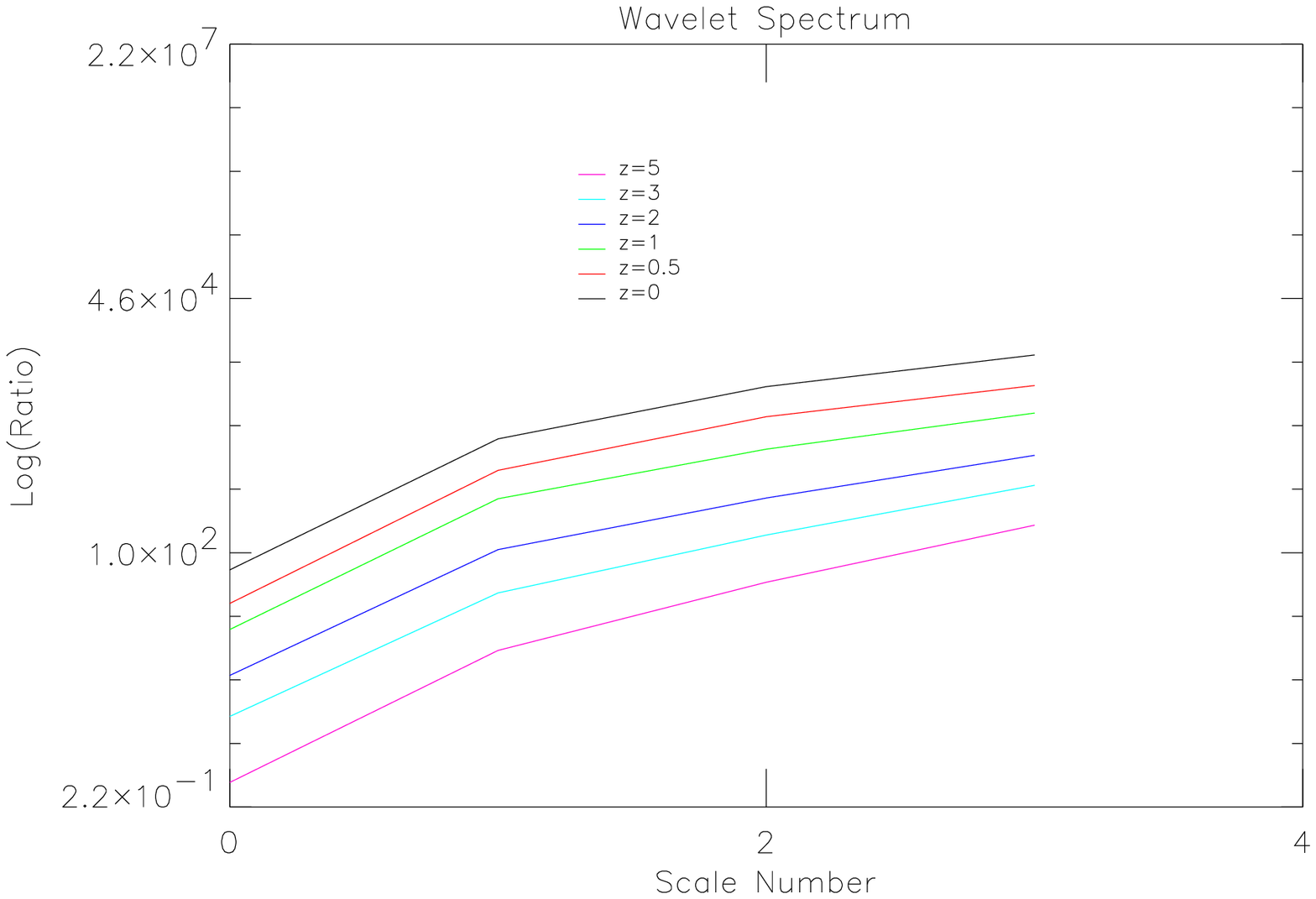,bbllx=1.2cm,bblly=12.5cm,bburx=19.5cm,bbury=25.5cm,width=15cm,height=7.cm,clip=}
}
\centerline{
\psfig{figure=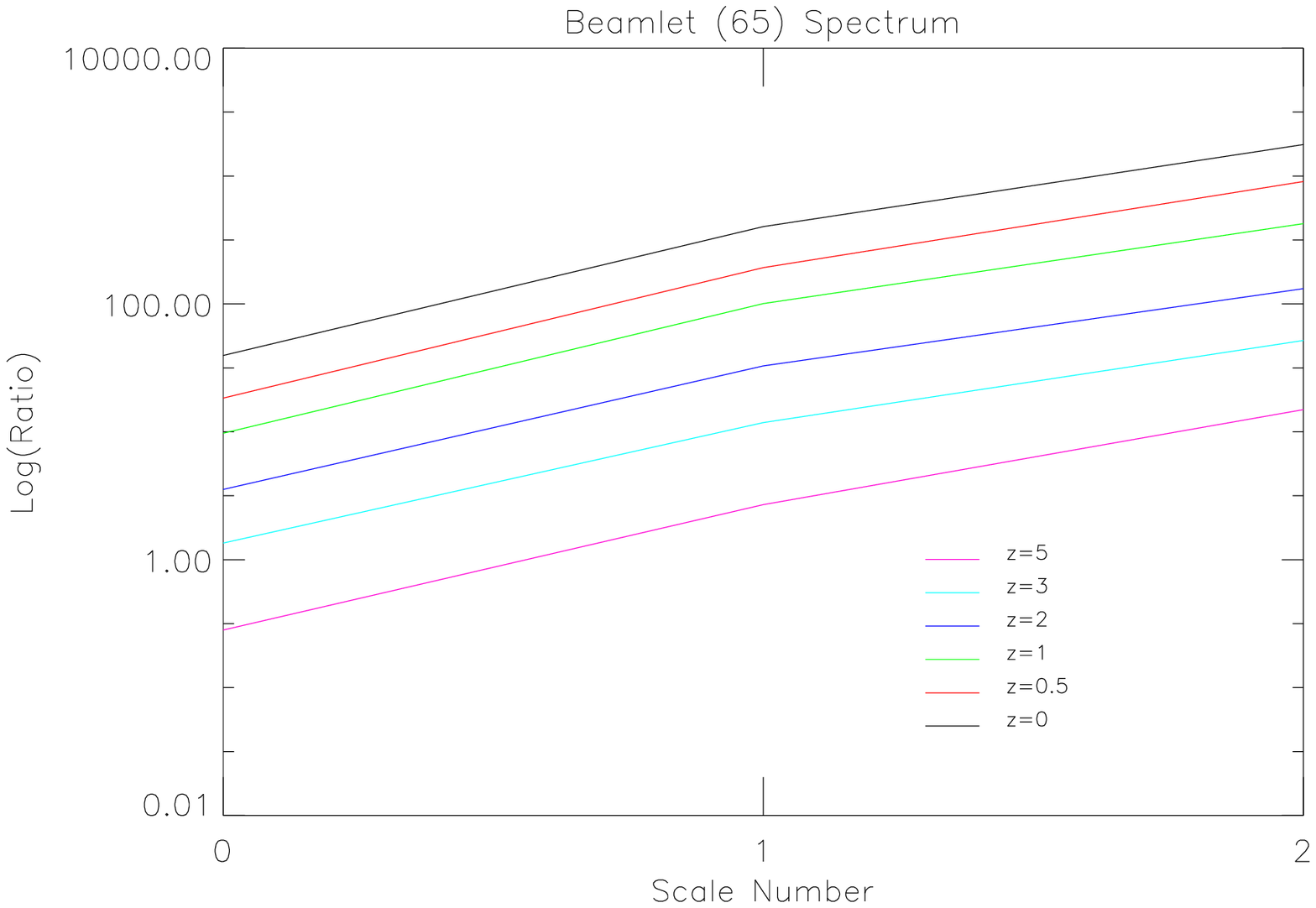,bbllx=1.2cm,bblly=12.5cm,bburx=19.5cm,bbury=25.5cm,width=15cm,height=7.cm,clip=}
}
\centerline{
\psfig{figure=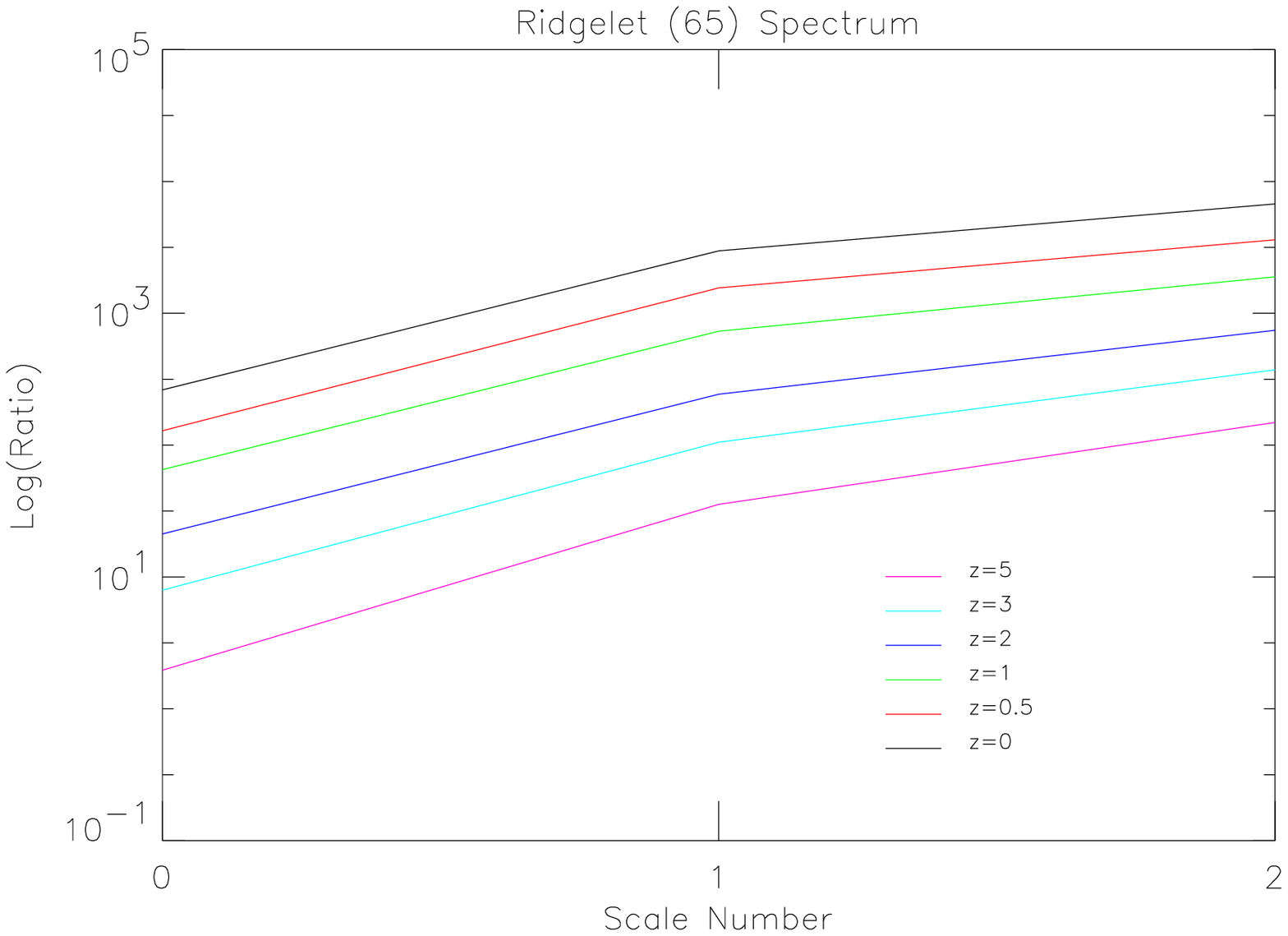,bbllx=1.2cm,bblly=12.5cm,bburx=19.5cm,bbury=25.5cm,width=15cm,height=7.cm,clip=}
}}
\caption{Top, Wavelet spectrum, middle, Beamlet spectrum, and bottom,
ridgelet spectrum at different redshifts.}
\label{fig_spectrum_sim_romain}
\end{figure}
Figure~\ref{fig_spectrum_sim_romain} shows respectively from
top to bottom the wavelet spectrum $P_w(z,j) = \sigma^2_{W,z,j}$,
the beamlet spectrum $P_b(z,j) = \sigma^2_{B,z,j}$ and
the ridgelet spectrum $P_r(z,j) = \sigma^2_{R,z,j}$.
In order to see the evolution of matter distribution
with the redshift and scale, we calculate the ratio
 $M_{bw}(j,z) = \frac{P_b(z,j)}{P_w(z,j)}$
and $M_{rw}(j,z) = \frac{P_r(z,j)}{P_w(z,j)}$.

\begin{figure}[htb]
\centerline{
\vbox{
\psfig{figure=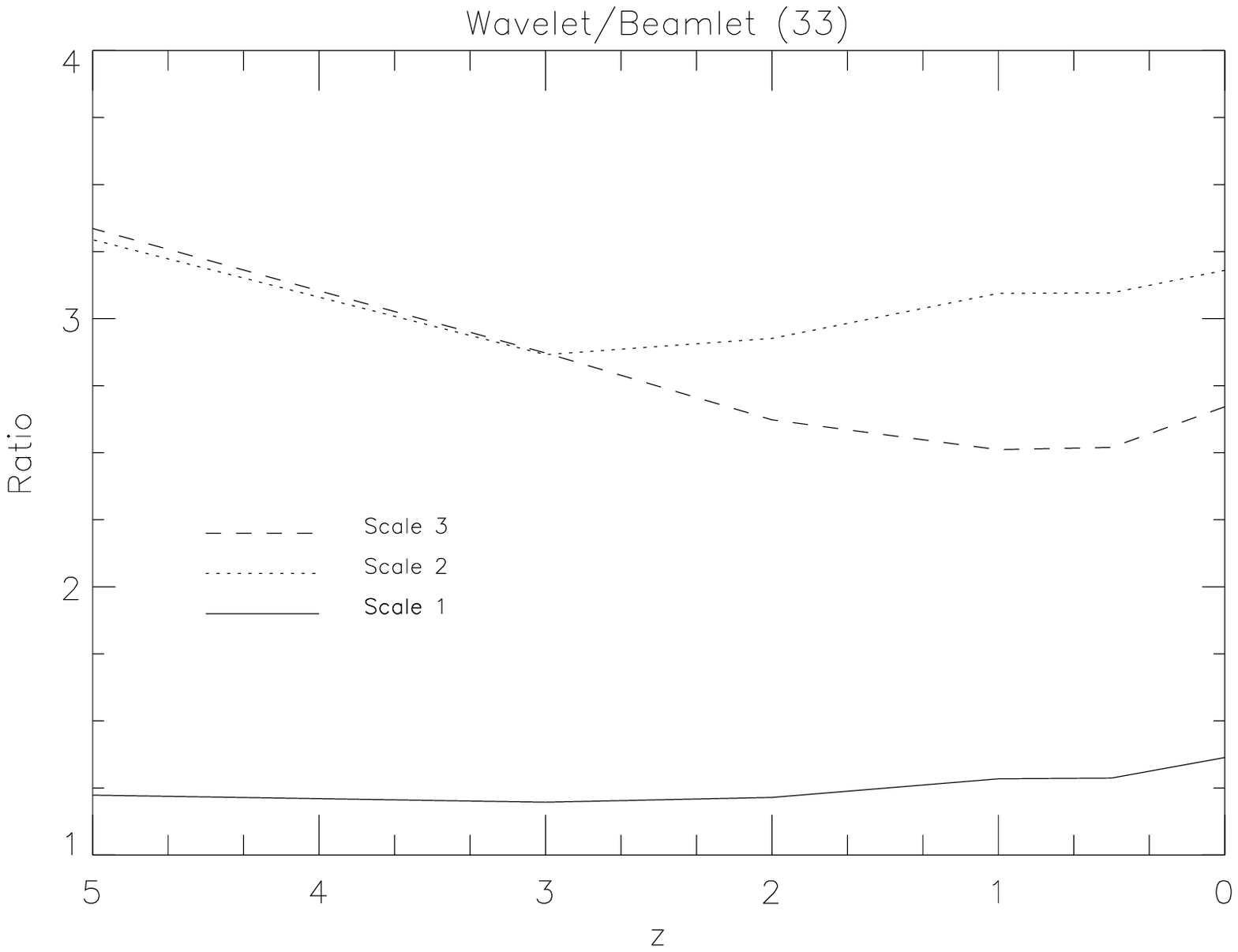,bbllx=1.5cm,bblly=12.5cm,bburx=19.5cm,bbury=25.5cm,width=15cm,height=9.cm,clip=}
\psfig{figure=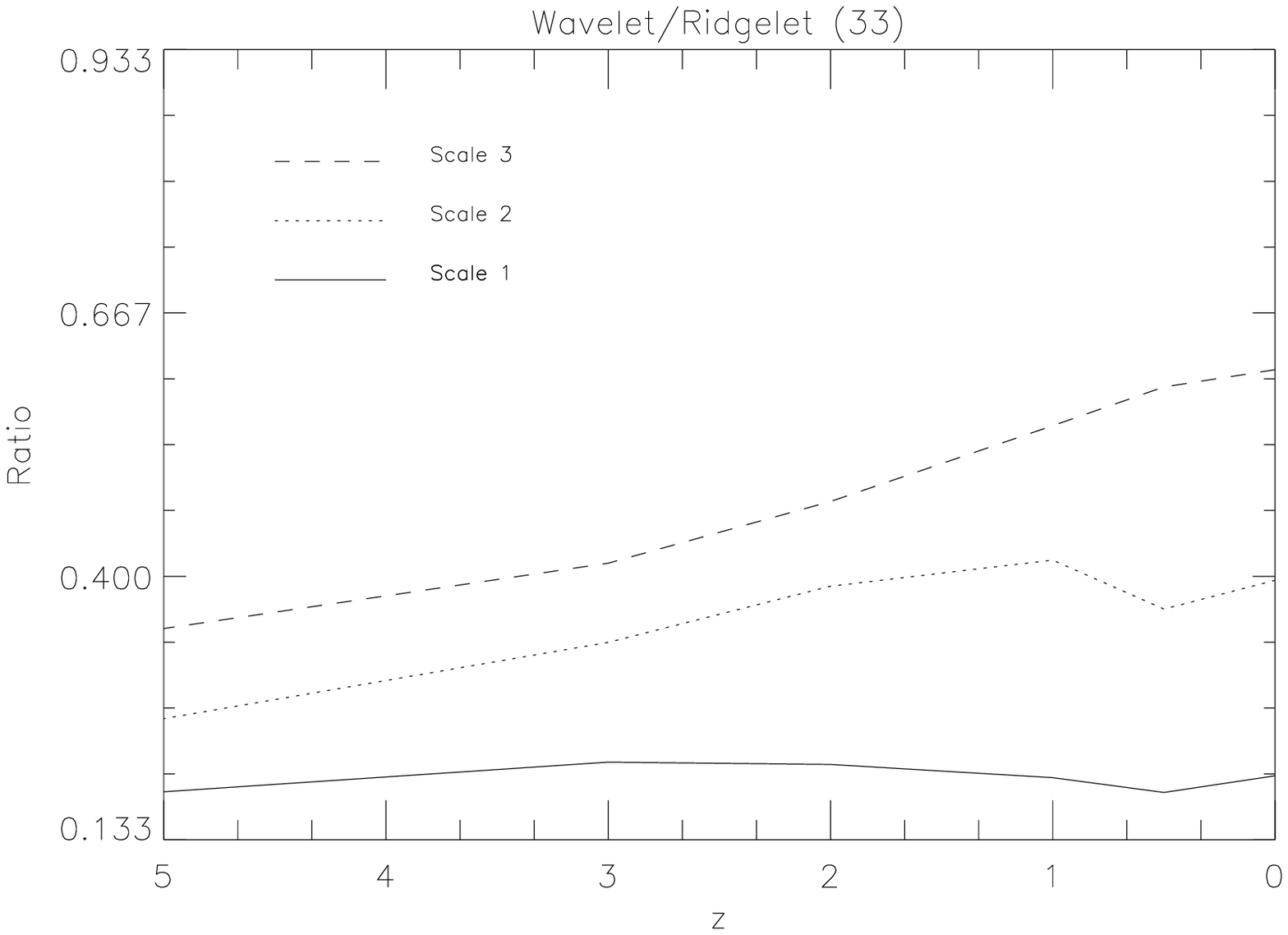,bbllx=1.5cm,bblly=12.5cm,bburx=19.5cm,bbury=25.5cm,width=15cm,height=9.cm,clip=}
}}
\caption{$M_1(z,j)$ (top) and $M_2(z,j)$ (bottom) for the scale number $j$ equals
to  1,2 and 3.}
\label{fig_ratio_sim_romain}
\end{figure}
\begin{figure}[htb]
\centerline{
\vbox{
\psfig{figure=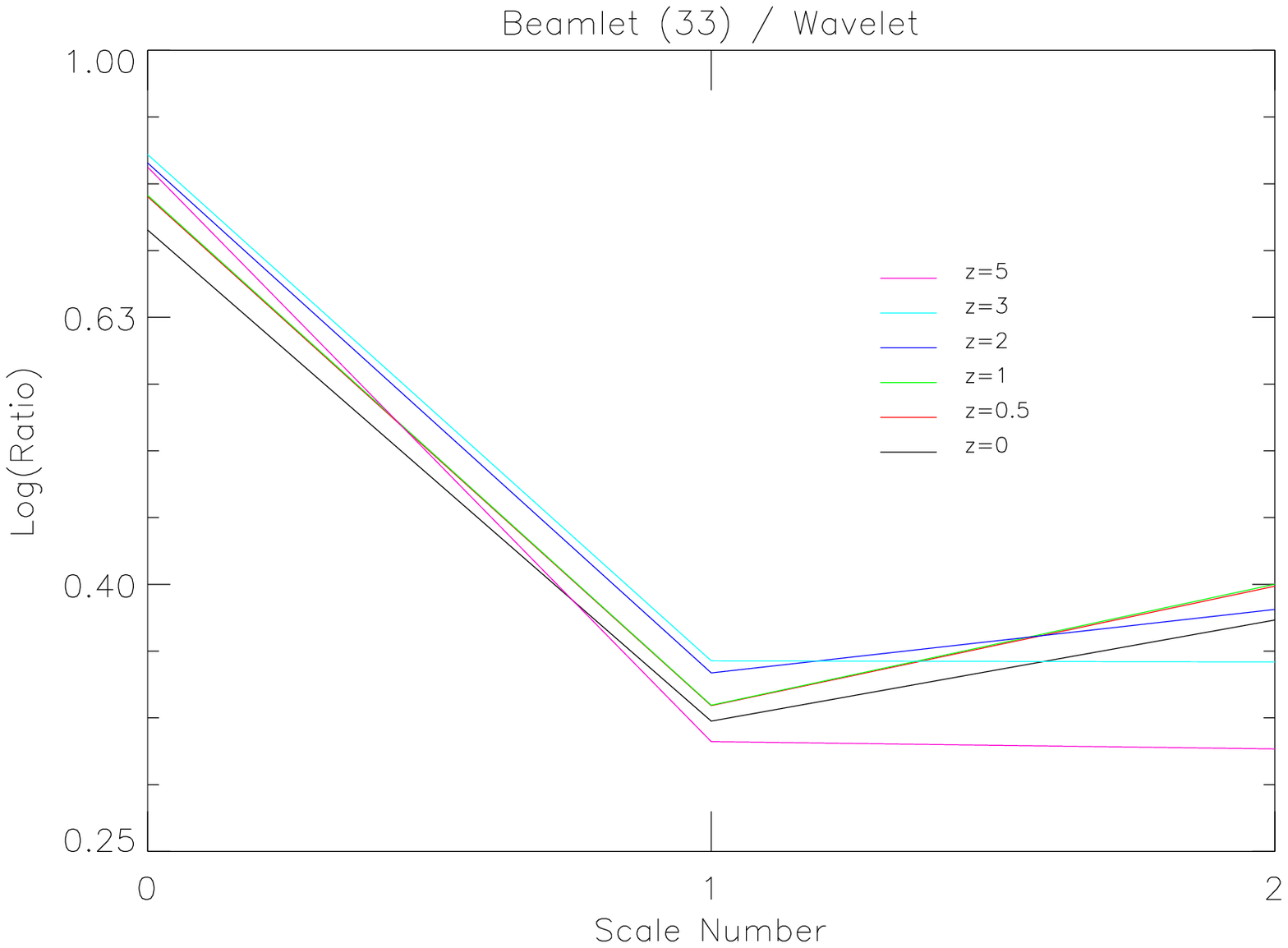,bbllx=1.5cm,bblly=12.5cm,bburx=19.5cm,bbury=25.5cm,width=15cm,height=9.cm,clip=}
\psfig{figure=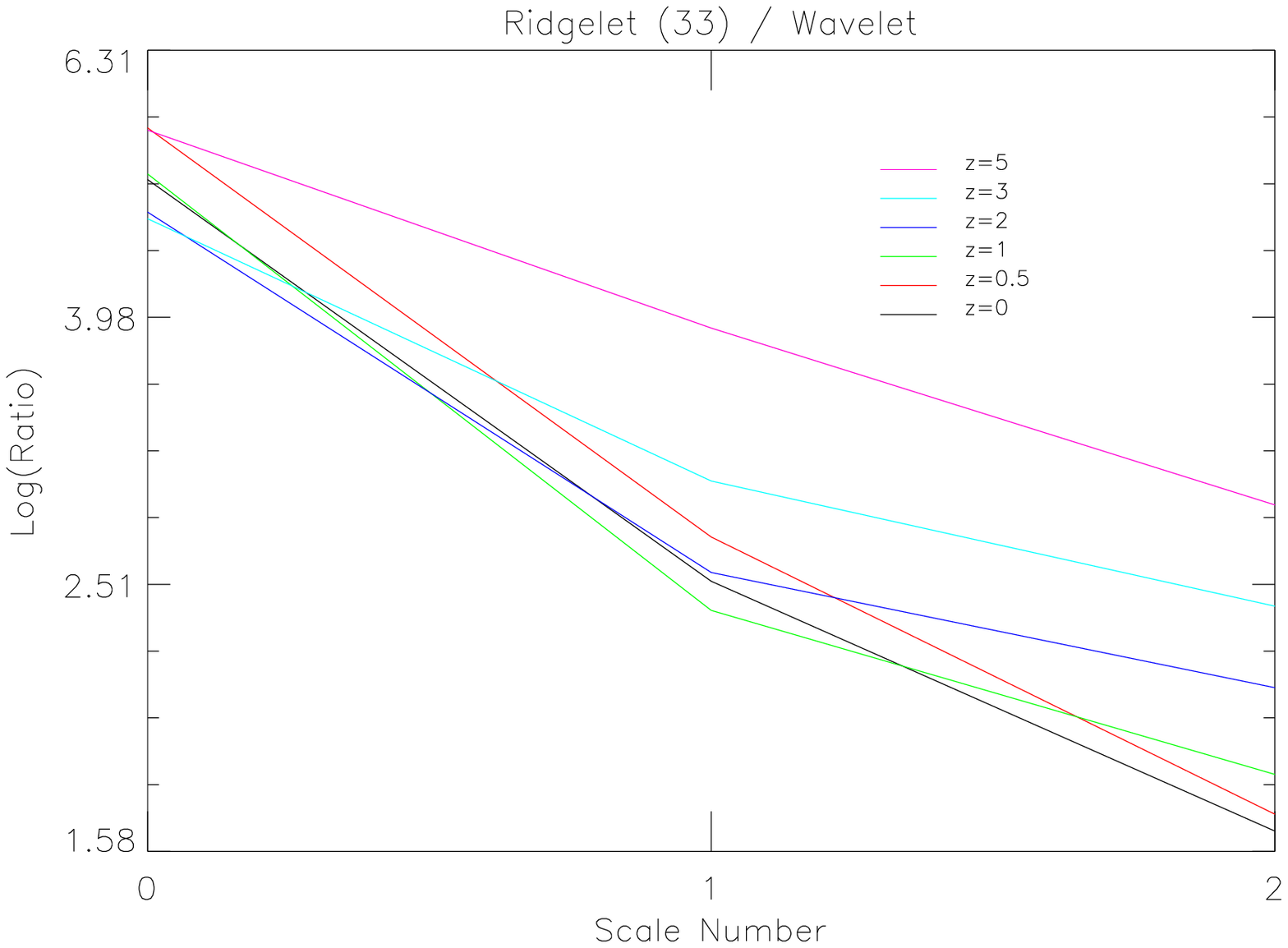,bbllx=1.5cm,bblly=12.5cm,bburx=19.5cm,bbury=25.5cm,width=15cm,height=9.cm,clip=}
}}
\caption{$M_1(z,j)$ (left) and $M_2(z,j)$ (right) at different redshifts.}
\label{fig_ratio_ps_sim_romain}
\end{figure}
Figure~\ref{fig_ratio_sim_romain} shows the $M_1$ and $M_2$ curves as a
function of $z$ and Figure~\ref{fig_ratio_ps_sim_romain}
shows the $M_1$ and $M_2$
curves as a function of the scale number $j$.

The $M_1$ curve does not show too much evolution, while the
$M_2$ curve presents a significant slope.
This shows that the beamlet transform
is much more sensible to the formation of clusters than
the ridgelet transform. This is not surprising since the
beamlet function support
is much smaller than the ridgelet function support. $M_2$ increases
with $z$ showing clearly the cluster formation.
This indicates that the combination of multiscale transformations
allows us to get some information about the degree of clustering,
filamentarity, and sheetedness.

\section{Conclusion}
We have introduced in this paper two new methods to analyze catalogs of galaxies.
The first one consists in estimating the real underlying density though a wavelet
denoising. Structures are first detected in the wavelet space, and an iterative reconstruction
is performed. A smoothness constraint based on the $l_1$ norm of the wavelet coefficients
is used, which reduce the amount of artifact in the reconstructed density, especially the ringing
artifacts around strong features which are due to the wavelet function shape.
We have shown that such an approach leads to much more reliable results than a Gaussian
filtering when we want to derive a Genus curve from the catalog. This could also be true for
other techniques which require to pre-process the data with a Gaussian filtering.
The wavelet denoising preserves the resolution of the detected features
whatever their sizes, and remove the noise in a non-linear way at all the scales.

The second approach does not require to detect anything. It is based on the analyzing of 
the distribution of coefficients obtained by several multiscale transforms.
We have introduced two new multiscale decompositions, the 3D ridgelet transform and the
3D beamlet transform. We have described how to implement them using the FFT. Then we have
shown that combining the information related to wavelet, ridgelet and beamlet coefficients leads
to a new way to describe a data set. We have used in this paper the skewness and the kurtosis, 
but other recent statistic estimator such the Higher Criticism \cite{gauss:lin02} could be used as well. 
Each multiscale transform is very sensible to one kind of feature, the wavelet to clusters, the 
beamlet to filaments, and the ridgelet to walls. A similar method has been proposed for analyzing
CMB maps \cite{starck:sta03_1} where both the curvelet and the wavelet transform
were used for the detection and the discrimination of non Gaussianities.
This combined multiscale statistic is very powerful 
and we have shown that two data set that cannot be distinguished using a two point correlation function
are clearly identified as different using our method. We believe that such an approach will permit to
better constraint the cosmological models. 

\section*{Acknowledgments}
We wish to thank Romain Teyssier for giving us the $\Lambda$-CDM simulated data used in the
third experiment. This work has been supported by the Spanish MCyT project
AYA2003-08739-C02-01 (including FEDER), by the Generalitat
Valenciana project GRUPOS03/170, and  by the National Science Foundation 
grant DMS-01-40587 (FRG),
and by the Estonian Science Foundation grant 4695.

\clearpage
\bibliographystyle{plain}
\bibliography{gauss,cf_ana,xwave,candes,entropy,starck,wave,restore,ima,astro,cluster,mc,curvelet}
\end{document}